\documentclass[preprint]{aastex}
\def\dm15{$\Delta$m$_{15}$($B$)}
\usepackage[footnotesize]{caption}

\begin{document}

\title{The Rise and Fall of Type Ia Supernova Light Curves in the SDSS-II Supernova Survey}

\author{Brian T. Hayden\altaffilmark{1},
Peter M. Garnavich\altaffilmark{1},
Richard Kessler\altaffilmark{2,3},
Joshua A. Frieman\altaffilmark{2,4,5},
Saurabh W. Jha\altaffilmark{6,7},
Bruce Bassett\altaffilmark{8,9}
David Cinabro\altaffilmark{10},
Benjamin Dilday\altaffilmark{7},
Daniel Kasen\altaffilmark{11,12},
John Marriner\altaffilmark{13},
Robert C. Nichol\altaffilmark{14},
Adam G. Riess\altaffilmark{15,16},
Masao Sako\altaffilmark{17},
Donald P. Schneider\altaffilmark{18},
Mathew Smith\altaffilmark{8},
Jesper Sollerman\altaffilmark{19,20}
} 
\altaffiltext{1}{University of Notre Dame, 225 Nieuwland Science Hall, Notre Dame, IN 46556, USA}
\altaffiltext{2}{Kavli Institute for Cosmological Physics, The University of Chicago, Chicago, IL 
60637, USA}
\altaffiltext{3}{Enrico Fermi Institute, The University of Chicago, Chicago, IL 60637, USA}
\altaffiltext{4}{Department of Astronomy and Physics, The University of Chicago, Chicago, IL 
60637, USA}
\altaffiltext{5}{Center for Particle Astrophysics, Fermi National Accelerator Laboratory, Batavia, IL 
60510, USA}
\altaffiltext{6}{Kavli Institute for Particle Astrophysics and Cosmology, Stanford University, Stanford, CA
94305-4060, USA}
\altaffiltext{7}{Department of Physics and Astronomy, Rutgers University, 136 Frelinghuysen Road, Piscataway, NJ 08854, USA}
\altaffiltext{8}{Department of Mathematics and Applied Mathematics, University of Cape Town, Rondebosch 7701, South Africa}
\altaffiltext{9}{South African Astronomical Observatory, Cape Town, South Africa}
\altaffiltext{10}{Department of Physics and Astronomy, Wayne State University, Detroit, MI, USA}
\altaffiltext{11}{Department of Astronomy and Astrophysics, University of California, Santa Cruz, CA 95064, USA}
\altaffiltext{12}{Hubble fellow}
\altaffiltext{13}{Center for Particle Astrophysics, Fermi National Accelerator Laboratory, P.O. Box 500, Batavia, IL 60510, USA}
\altaffiltext{14}{Institute of Cosmology \& Gravitation, University of Portsmouth, Portsmouth PO1 3FX, UK}
\altaffiltext{15}{Space Telescope Science Institute, 3700 San Martin Drive, Baltimore, MD 21218, USA}
\altaffiltext{16}{Department of Physics and Astronomy, Johns Hopkins University, 3400 North Charles Street, Baltimore, MD 21218, USA}
\altaffiltext{17}{Department of Physics and Astronomy, University of Pennsylvania, 209 South 33rd Street , Philadelphia, PA 19104, USA}
\altaffiltext{18}{Department of Astronomy and Astrophysics, The Pennsylvania State University, 525 Davey Laboratory, University Park, PA 16802, USA}
\altaffiltext{19}{Dark cosmology centre, Niels Bohr Institute, University of Copenhagen, Juliane Maries Vej 30, DK-2100 Copenhagen O, Denmark}
\altaffiltext{20}{Department of Astronomy, The Oskar Klein Centre, Stockholm University, 106 91 Stockholm, Sweden}

\begin{abstract}
We analyze the rise and fall times of type~Ia supernova (SN~Ia) light curves discovered by the 
Sloan Digital Sky Survey-II (SDSS-II) Supernova Survey. From a set of 391 light curves $k$-corrected to the rest-frame $B$ 
and $V$ bands, we find a smaller dispersion in the rising portion of the light curve compared to the 
decline. This is in qualitative agreement with computer models which predict that variations in 
radioactive nickel yield have less impact on the rise than on the spread of the decline rates. The
differences we find in the rise and fall properties suggest that a single `stretch' correction to the
light curve phase does not properly model the range of SN~Ia light curve shapes.
We select a subset of 105 light curves well observed in both rise and fall portions of the light 
curves and develop a `2-stretch' fit algorithm which estimates the rise and fall times independently. 
We find the average time from explosion to $B$-band peak brightness is $17.38\pm 0.17$ days, but 
with a spread of rise times which range from 13 days to 23 days. Our average rise time is shorter 
than the $19.5$ days found in previous studies;  this reflects both the different light curve template 
used and the application of the 2-stretch algorithm. The SDSS-II supernova set and the local SNe~Ia 
with well-observed early light curves show no significant differences in their average rise-time 
properties. We find that slow-declining events tend to have fast rise times, but that the distribution of 
rise minus fall time is broad and single peaked. This distribution is in contrast to the bimodality in this 
parameter that was first suggested by \citet{str07} from an analysis of a small set of local SNe~Ia. We 
divide the SDSS-II sample in half based on the rise minus fall value, $t_r-t_f \leq 2$ days  and $t_r-
t_f>2$ days, to search for differences in their host galaxy properties and Hubble residuals; we find 
no difference in host galaxy properties or Hubble residuals in our sample. 

\end{abstract}


\section{Introduction}

Type~Ia supernovae (SNe~Ia) are bright stellar explosions important for their role as distance 
indicators. SN~Ia distances have been used to constrain the value of the Hubble constant 
\citep{jha99, freed01}, and they also showed that our universe has a lower than critical matter  
density \citep{gar98, perl98}. \citet{rie98}  and \citet{perl99} used distant Ia SNe to show that the
universe currently has an accelerating rate of expansion implying a `dark' energy component. As
more SN~Ia observations have confirmed this result \citep{knop03, ton03, bar04, rie04}, the focus 
has shifted to constraining the properties of dark energy \citep{gar98b, ast06, rie07, wv07, mik07, 
eis07}. Systematic uncertainties in measuring dark energy parameters with supernovae 
now dominate over statistical errors \citep{kessler08} and a better understanding of supernova 
physics may help to constrain dark energy properties using SNe~Ia.

It is widely accepted that SNe~Ia are the result of thermonuclear explosions of carbon-oxygen white
dwarfs (WDs), but the nature of the progenitor remains uncertain. In most models, the explosion 
occurs when the WD nears the Chandrasekhar limit by gaining mass from a binary 
companion. This mass gain is achieved either by single-degenerate (SD) mass transfer or WD 
coalescence in a double-degenerate (DD) scenario. For a comprehensive review of these models, 
see \citet{liv00}. Observers have attempted to distinguish between the two models, with conflicting 
results. \citet{how06} found that the extremely luminous supernova SNLS-03D3bb (SN~2003fg) 
had low ejecta velocity that could have resulted from a super-Chandrasekhar progenitor. The high 
total mass and large nickel yield suggested that SNLS-03D3bb was the product of a DD merger. \citet{hic07} found that SN 2006gz had attributes consistent with the DD model. 
Its spectrum showed significant amounts of unburned carbon, and a low silicon velocity at early 
phases.  The very broad light curve implies a large yield of radioactive nickel as expected in the DD 
model \citep{hic07}, although late-time observations show only weak iron emission lines
\citep{maeda08}. Even though the SD model is generally accepted as the most plausible
\citep{liv00}, there is evidence that it might not be the complete story \citep{hic07,pri08}. Some 
mixture of progenitors may be producing events with subtle differences in character. The explosion 
rates for these progenitor systems will depend on the age and the star formation history of the 
parent population, and the ratio of the different progenitors may not be constant through the history 
of the universe \citep{ham96,gal05,sul06,man06,gal08}.

Understanding the origin of the Phillips relation (the correlation between light curve shape and peak 
luminosity; \citet{phil93, kasen07})  and its dispersion is critical to improving SNe~Ia as reliable distance 
indicators. \citet{tim03} argued that radioactive nickel yield should be determined by progenitor 
metallicity, but age of the host galaxy stellar population appears to control the mass of $^{56}$Ni 
\citep{ham96,gal05,sul06}. Progenitor metallicity may play a secondary role \citep{gal08} or have no 
significant effect on radioactive yield \citep{how08}. Recent models by \citet{woos07} show that 
variations in kinetic energy (KE), metallicity, and mixing between burning layers provide light curves beyond 
the range of the observed Phillips relation, implying that not all combinations of variables are found in 
real SNe~Ia.

SN~Ia luminosity and light curve shape may be influenced by the physics of the explosions. There is
a consensus that normal SNe~Ia result from a detonation (supersonic burning) of much of the progenitor 
WD, but it likely begins as a deflagration (subsonic fusion front) to allow expansion and some 
burning at low densities. This model was first introduced by \citet{khok91}; \citet{hoef96} showed that the 
model matched the observed light curves and spectra very well. In the ``delayed detonation" scenario, 
the timing of the transition from deflagration to detonation provides a natural means of varying the 
radioactive nickel yield and generating the observed range of luminosities and decline rates 
\citep{arnett94}. Deflagrations tend to be very asymmetric, and if the asymmetries survive the detonation 
then viewing angle can result in perceived variations even for similar explosions \citep{kasenplewa07}.

Study of the early-time light curve may be important in diagnosing the progenitor problem and explosion 
physics \citep{hoef96}. The work by \citet{rie99} to constrain the early light curves of local supernovae 
established the decline-rate-corrected average rise time of 19.5 $\pm$ 0.2 days. Here, ``rise time" is 
defined as the time elapsed from explosion to peak \textit{B}-band flux. \citet{ald00} and \citet{gold01} 
demonstrated the consistency between high- and low-redshift rise times, and similar results were found 
by \citet{con06} with a large set of SNLS \citep{pri05} supernovae. \citet{garg07} studied the rise times of 
SNe~Ia discovered behind the Large Magellanic Cloud during the SuperMACHO survey lens project 
and measured a 17.6$\pm 1.3$-day rise time in the equivalent of the $V$ bandpass.

Recently, \citet{str07} analyzed eight low-redshift SNe~Ia with well-observed early light curves from a 
new perspective. A single ``stretch" parameter has commonly been used to describe the full range 
of $B$ and $V$ light curves by compressing or expanding the time axis around the epoch of peak 
brightness \citep{perl97}. \citet{str07} decoupled the rise and decline portions of the eight nearby 
light curves and  found that SNe~Ia can have a range of rise times for a given decline rate. His rise-time 
minus fall-time distribution ($t_r-t_f$; hereafter defined as RMF) roughly divided his small supernova 
sample into two groups, possibly suggesting two progenitors or  explosion mechanisms. A larger set of 
SNe is required to rigorously test this possibility.
 
Here, we analyze light curves from the Sloan Digital Sky Survey-II (SDSS-II) Supernova
Survey \citep{fri08} which spectroscopically identified approximately 500 SNe~Ia over its three-year 
lifetime. The SDSS-II Supernova Survey scanned 300 deg$^2$ of sky with a cadence 
as rapid as 2~days between visits (weather and lunar phase often increased the time between 
observations of the same field), making the survey well suited to an early rise-time study. We also 
introduce a new fitting method that independently estimates the rise and the fall times of SNe~Ia light 
curves.

\section{Data}

\subsection{SDSS-II Light Curves}

The SDSS-II Supernova Survey \citep{fri08} was designed to find and 
characterize several hundred SNe~Ia at intermediate redshifts in order to fill in the supernova 
``desert" between the nearby discoveries and the``high-$z$" events. Mapping the expansion history 
at $z\approx 0.2$ provides unique tests of cosmological models and constrains systematic errors 
\citep{kessler08,davis08,lamp08}. The SDSS-II supernovae also provide a large, uniform, high-quality
sample of SNe~Ia to study the properties of these explosions. The SDSS-II Supernova Survey 
operated three campaigns between 2005, 2006 and 2007 September and December  using the 
2.5 m telescope \citep{gunn06} at Apache Point Observatory to scan 300 deg$^2$ of 
sky as often as every-second night. Template images were subtracted from each new night of data 
and software scanned for new variable objects. Every candidate supernova was inspected 
visually to avoid image artifacts and asteroids. Transients with a high probability of being SNe~Ia 
\citep{sako08} were queued for spectrographic observation \citep{zheng08}. The SDSS-II SN 
survey takes advantage of the extensive database of reference images, object catalogs, and 
photometric calibration compiled by the SDSS (see \citet{york00} for an overview of the SDSS). The 
early data release of the SDSS can be found in \citet{stoughton02}. Data release seven, the final 
data release of SDSS-II, can be found in \citet{aba09}.

The supernova sample minimizes photometric errors by using a single filter set and a photometric 
system \citep{smith02} calibrated by the original SDSS project \citep{fuk96,gunn98}. The 
photometric quality is assessed in \citet{ivez04}, and the astrometric calibrations are described in 
\citet{pier03}. Specifically, the area of the SDSS-II SN survey has been calibrated to 1\% 
photometric error \citep{smith02}. The photometric calibration employed several minor telescopes 
as well; this Monitor Telescope Pipeline (MTPIPE) is described in \citet{tuck06}. 

A software ``robot'' was developed that automatically reduced CCD observations of standard stars 
in order to assess photometricity and build data on site conditions at the Apache Point Observatory 
\citep{hogg01}. Using this tool to select the best observing nights, weather and lunar phase extend 
the time between observations to an average of about 4.5 days \citep{fri08}. The rapid cadence of 
the SDSS-II SN survey assures that most of the SDSS SNe have well-defined, densely sampled 
light curves even on the rise portion. The data include pre-explosion flux measurements, so that 
many of the SDSS-II Supernova Survey type Ia light curves are among the most well-sampled 
and well-calibrated light curves that have been studied to date.

The SDSS-II SN survey identified 498 spectroscopically confirmed SNe Ia (number 
accurate to the time of our analysis). Five of these are extremely peculiar and are not included in this 
study. SN 2007qd \citep{colin} and SN 2005hk \citep{phil07} are categorized as 2002cx-like objects. SN 
2002cx showed evidence for a very low expansion velocity compared to `normal' SNe~Ia as well as an 
unusually low peak luminosity, leading to the hypothesis that it was a pure deflagration event 
\citep{branch04, jha06}. SN 2005gj, SN 7017, and SN 15557 \citep{pri07} are 2002ic like. SN~2002ic 
showed a type~Ia spectrum along with strong hydrogen emission lines \citep{hamuy03} suggesting a 
shock interacting with circumstellar material.
% Lastly, SN 2005hj \citep{quimby07} shows an expansion velocity plateau similar 
%to SN1991T or SN1999aa; these observations in concert with current models suggest that 
%SN2005hj may be physically distinguishable from other normal-bright SN Ia. **** 05hj=SN6558 and is in the sample!!!

We limit the events to redshift less than 0.4 and require that there be at least one photometric epoch 
more than 2 days before maximum and more than 5 days after maximum. These requirements 
result in a sample of 391 supernovae. The photometry is measured using ``scene modeling" developed 
by \citet{holtz08}. The Sloan $g$, $r$, and $i$ photometry has been $k$-corrected to Bessell $B$ and $V
$ bands using SNANA \citep{kess09b,hsiao07}. We use the time of maximum obtained 
by the fitting procedure in SNANA as the initial value for our $\chi^2$ minimizing function.  The 
redshift of the 391 SNe~Ia ranges from 0.037 to 0.40 with a median redshift of 0.21.

\begin{figure}[h!]
\begin{tabular}{cc}
\includegraphics*[scale=0.5]{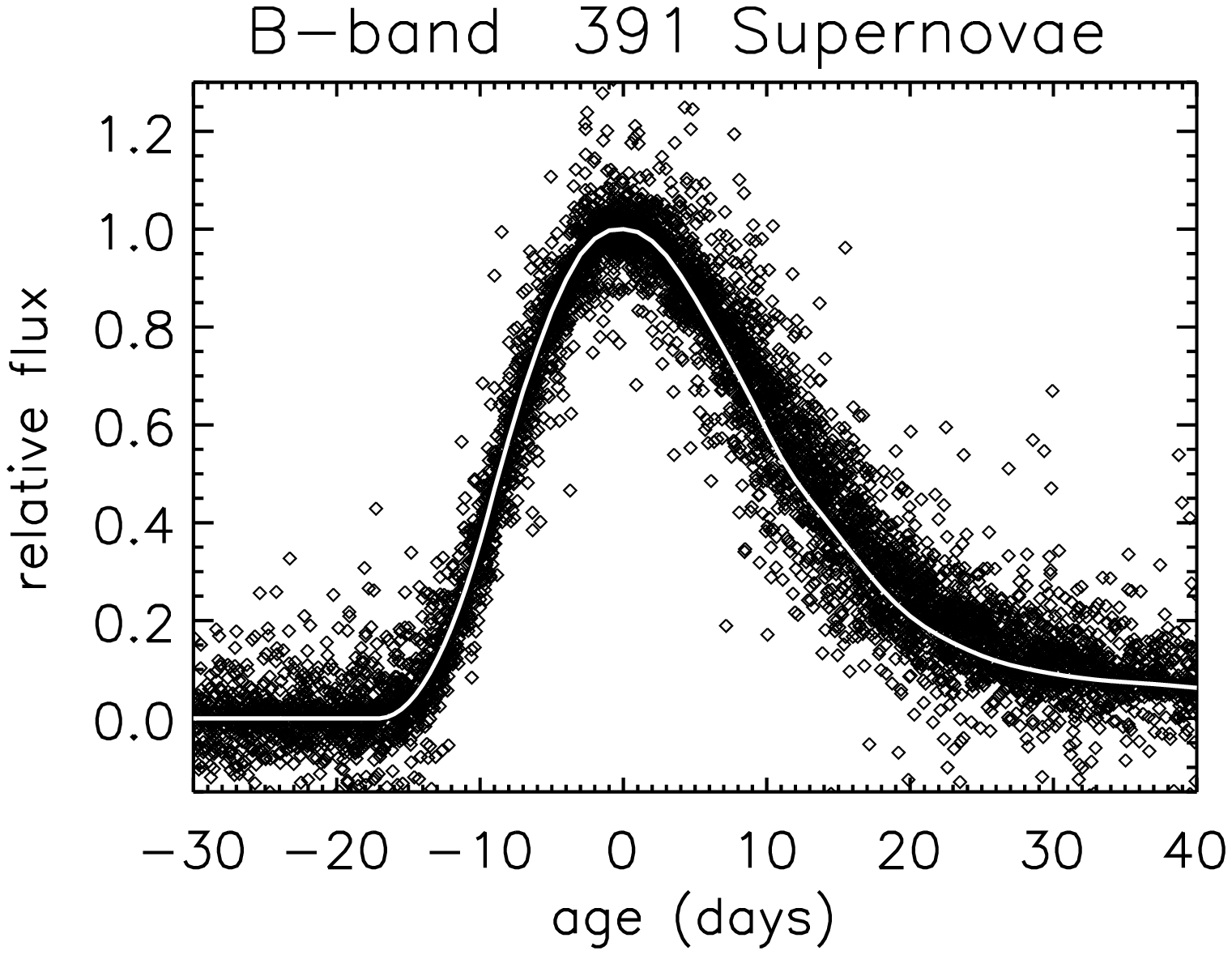} & \includegraphics*[scale=0.5]{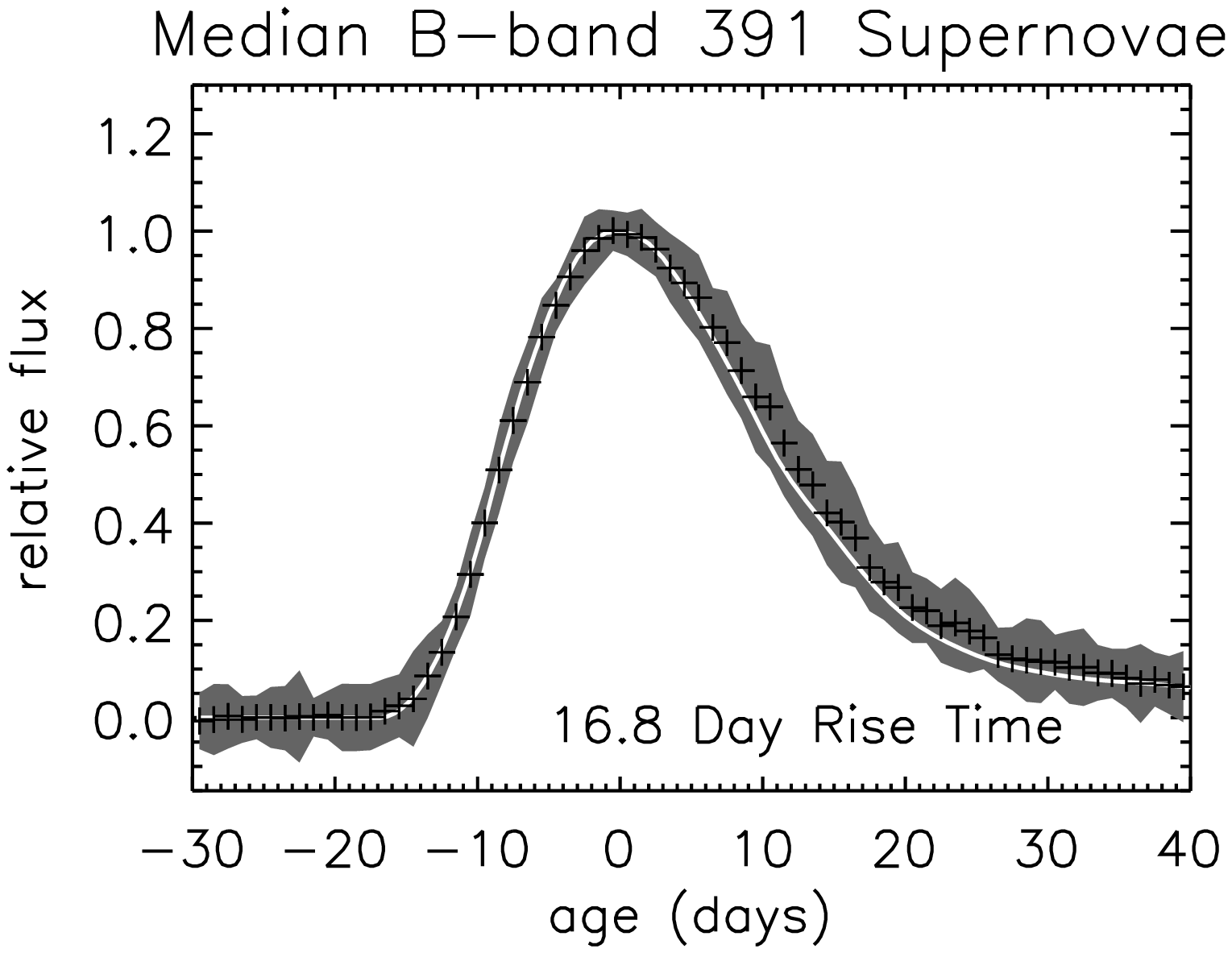}\\
\includegraphics*[scale=0.5]{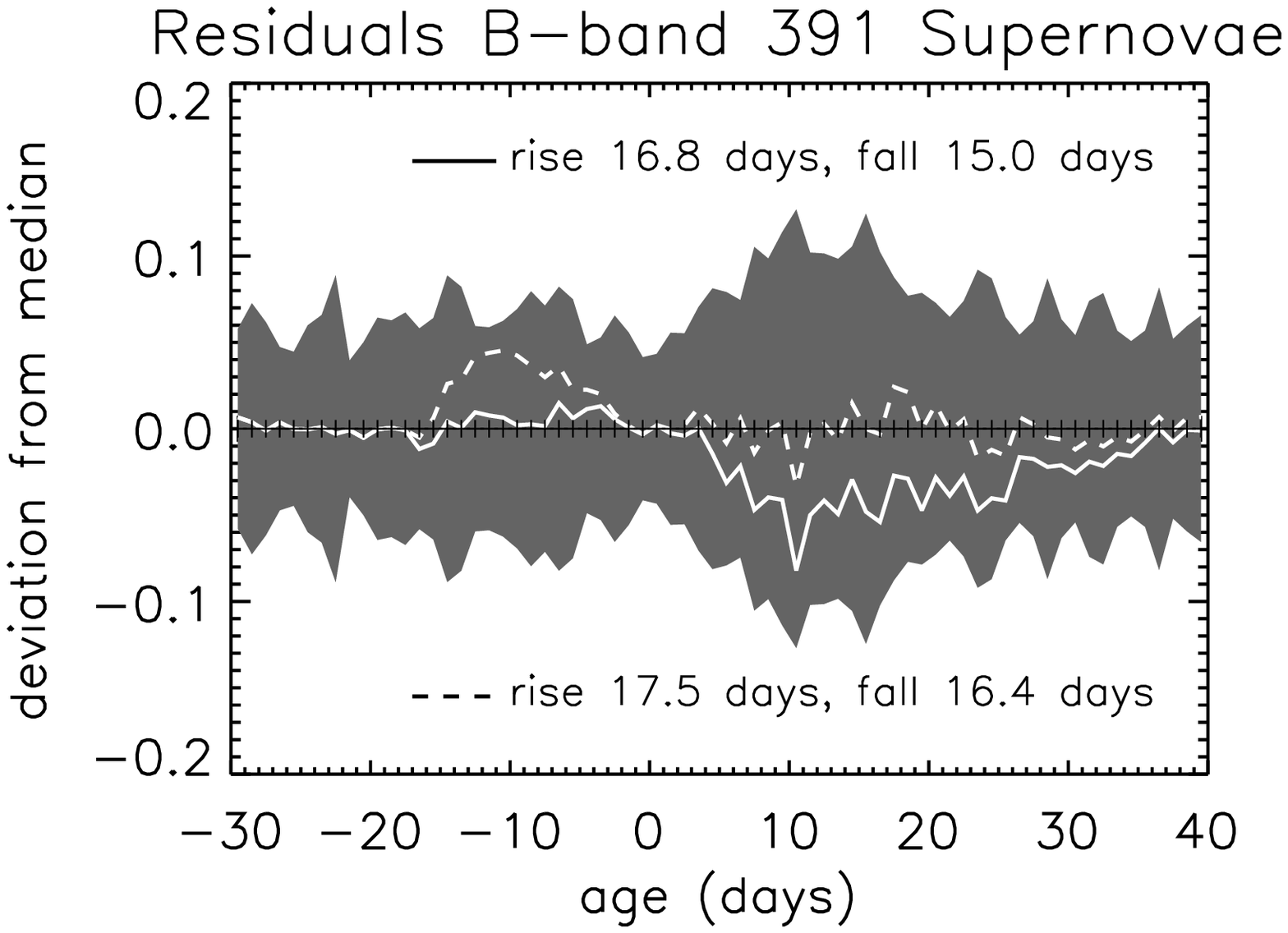}
\end{tabular}
\caption{{\it Top left:} individual SDSS-II supernova observations plotted in the rest frame and 
normalized to a peak flux of unity. The light curves were $k$-corrected to $B$ band using SNANA. 
Overplotted are the MLCS2k2 fiducial light curves with a 16.8-day rise time. {\it Top right:} median 
of all the data points binned by day are plotted (+) and the gray regions show the rms deviations 
about the median values. The white lines show the MLCS2k2 fiducial curves.  {\it Bottom:} fluctuations 
about the median are plotted along with template light curve residuals. This clearly shows the scatter in 
rise time is significantly less than in the decline portion of the light curve. The solid line shows the 
MLCS2k2 template with a 16.8-day rise time is an excellent match to the median of the 391 SNe Ia. The 
dashed line indicates that increasing the rise time by 0.7 days is clearly a poor fit to the median rise data 
while a decline of 16.4~days is a good match.
\label{alldata}}
\end{figure}
\clearpage

Figure \ref{alldata} displays the entire set of 391 $B$ and $V$ light curves after estimating the time 
of maximum, normalizing the peak to unity, and correcting for time dilation. 
%Also shown is one of the MLCS2k2 fiducial light curves; it is clear that it matches the shape of the 
%real data very well. 
The data were binned in 1-day intervals and the median value calculated for each day, also 
shown in Figure \ref{alldata}. The MLCS2k2 fiducials are an excellent match to the median light 
curves, with an extrapolation of 16.8 days. The fading portion of the SDSS-II $B$-band data is brighter 
than the fiducial, indicating that the supernovae discovered by SDSS-II are, on average, slower than 
typical low-redshift events. For the $V$ band, the fiducial and the SDSS-II light curves match well. It is 
clear that the scatter about the median light curve is smaller during the rise than during the decline. The 
maximum root-mean-square (rms) dispersion for $B$ band in 1-day bins is 0.09 (scaled flux units) 
on the rise and 0.13 on the decline. The average rms about the median on the rise is 0.069 while on 
the fall it is 0.091. This implies a smaller range of rise times than fall times and therefore that the full 
range of SN~Ia light curve shapes cannot be well fitted by a single-stretch parameter. 

\citet{kasen09} has provided a theoretical prediction of the impact on the light curve of SN Ia due to the 
interaction of the WD explosion with a potential companion star. It is estimated that approximately 10\% 
of SN Ia explosions that occur in the SD channel would have light curves that are 
affected by the shock interaction of the explosion with a binary companion. This effect is not immediately 
obvious in the SDSS-II from any analysis in this paper; however, an in-depth statistical study is in 
progress to rigorously test the SDSS-II SN Ia sample for this effect \citep{hay09}.

\subsection{The Low-Redshift SN Ia Set}

In order to compare our fitter with the results in \citet{str07}, we applied our 2-stretch fitting method
to the same set of eight low-$z$ SNe~Ia studied by Strovink. These supernovae are SN~1990N 
\citep{90N}, SN~1994D \citep{94D}, SN~1998aq \citep{98aq}, SN~2001el \citep{01el}, 
SN~2002bo \citep{02bo}, SN~2003du \citep{03du}, SN~2004eo \citep{04eo}, and SN~2005cf 
\citep{05cf}. \citet{str07} combines light curve observations from different studies in his analysis of 
low-$z$ supernovae, but this has the danger of conflicting calibrations and differences in photometric 
error estimates. Instead, we limit observations to a single published data set selected based on 
cadence and number of data points before maximum.  Table \ref{tbl-1} compares the Strovink and 
the 2-stretch fitter results. We find very similar values for these supernovae.  The average 
difference between the rise time measured by \citet{str07} and our 2-stretch fitter is $-0.23 \pm 
0.58$ days (standard deviation) and the difference between fall times is $0.15 \pm 0.47$ days (standard 
deviation).

\begin{deluxetable}{lccccccc}
\tabletypesize{\scriptsize}
\tablecaption{Comparison of 2-stretch Method versus \citet{str07}\label{tbl-1}}
\tablewidth{0pt}
\tablehead{
\colhead{SN} &
\colhead{$\Delta m_{15}$\newline (mag)} & \colhead{$\Delta m_{15}$ (mag)} &
\colhead{$t_{rise}$ (days)} & \colhead{$t_{rise}$ (days)} &
\colhead{$t_{r} - t_{f}$ (days)} & \colhead{$t_{r}- t_{f}$ (days)} \\

\colhead{ } &
\colhead{Strovink07} & \colhead{2-Stretch} & \colhead{Strovink07} & \colhead{2-Stretch} &
\colhead{Strovink07} & \colhead{2-Stretch}
}
\startdata
SN 1990N  & 0.990 $\pm$ 0.034 & 1.029 $\pm$ 0.017 & 20.01 $\pm$ 0.46 & 19.81 $\pm$ 0.25 & 4.04 $\pm$ 0.74 & 4.31 $\pm$ 0.26\\  
SN 1994D  & 1.344 $\pm$ 0.021 & 1.422 $\pm$ 0.023 &  15.39 $\pm$ 0.47 & 15.43 $\pm$ 0.23 & 2.04 $\pm$ 0.53 & 2.73 $\pm$ 0.23\\ 
SN 1998aq & 1.042 $\pm$ 0.021 & 1.089 $\pm$ 0.013 & 17.52 $\pm$ 0.58 & 17.13 $\pm$ 0.22 & 2.03 $\pm$ 0.64 & 2.20 $\pm$ 0.22\\ 
SN 2001el  & 1.168 $\pm$ 0.021 & 1.080 $\pm$ 0.010 & 18.00 $\pm$ 0.56 & 18.22 $\pm$ 0.18 & 3.57 $\pm$ 0.63 & 3.21 $\pm$ 0.18\\ 
SN 2002bo  & 1.162 $\pm$ 0.031 & 1.168  $\pm$ 0.031 & 16.05 $\pm$ 0.37 & 17.50 $\pm$ 0.19 & 1.57 $\pm$ 0.53 & 3.21 $\pm$ 0.21\\ 
SN 2003du  & 0.982 $\pm$ 0.021 & 1.013 $\pm$ 0.010 & 17.71 $\pm$ 0.35 & 17.77 $\pm$ 0.17 & 1.64 $\pm$ 0.44 & 2.11 $\pm$ 0.17\\ 
SN 2004eo  & 1.403 $\pm$ 0.037 & 1.326 $\pm$ 0.010 & 16.64 $\pm$ 0.44 & 16.62 $\pm$ 0.13 & 3.86 $\pm$ 0.53 & 3.39 $\pm$ 0.13\\ 
SN 2005cf  & 1.068 $\pm$ 0.021 & 1.045 $\pm$ 0.008 & 16.62 $\pm$ 0.25 & 17.31 $\pm$ 0.10 & 1.35$\pm$ 0.36 & 1.97 $\pm$ 0.10\\ 
\enddata
%\tablecomments{A comparison of \citet{str07}'s numbers for the fit to the low-z supernovae. }
\end{deluxetable}

\section{2-Stretch Fitting Method}

With a handful of well-observed events, \citet{str07} found that the rise time of SNe~Ia can vary for a 
fixed decay time. He developed a fitting method called Adaptive QUArtic Algorithm (AQUAA)
which employs quartic splines with fuzzy knots to create smooth $UBVRI$ light curves for individual 
supernovae. The \citet{str07} fitter is computer intensive and does not work as well with noisy data; 
both are drawbacks when dealing with the large quantity and diverse quality of SDSS-II 
supernova data (see Section 2.1).

For the SDSS-II data, we desired a fitting method which maintains the simplicity of fitting light 
curves with the single-stretch technique, while decoupling the rise and fall portions during the 
fitting process.  Our `2-stretch' method attempts to accomplish this goal by stretching the pre-maximum 
portion separately from the post-maximum portion of the light curve. Our template curve is required to 
stay continuous at the joining point: peak brightness. Maximum light has a zero first derivative, making 
this a natural cutting point for using a 2-stretch method. A $\chi^2$ minimization is performed to 
determine the four parameters of the fit: rise stretch $(s_r)$, fall stretch $(s_f)$, time of maximum, and 
peak flux. It is not possible to completely remove the covariance between rise and fall times, because 
photometric errors and sampling gaps produce uncertainty in the time of maximum which mixes rise and 
fall times. Fitting simulated light curves with known rise and fall times allows us to estimate the 
correlation between rise and fall errors.

\begin{figure}[h!]
\begin{center}
\includegraphics*[scale=0.6]{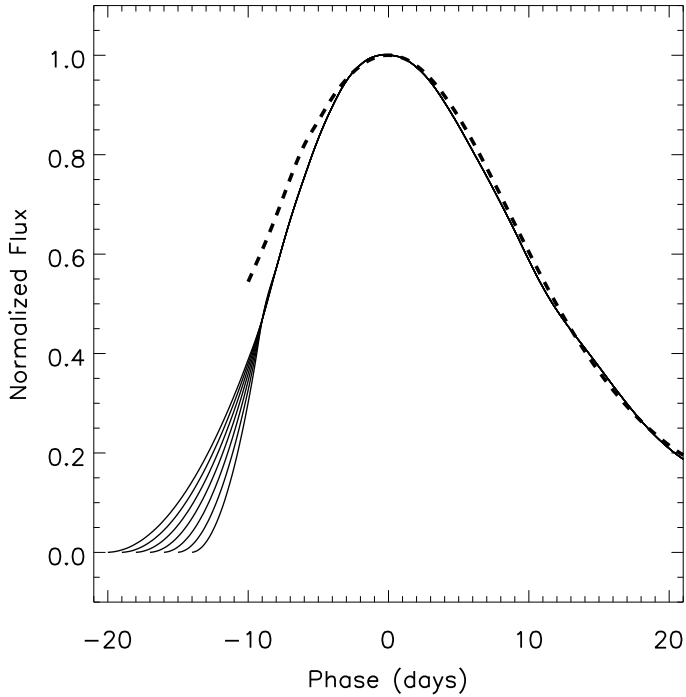}
\end{center}
\caption{Comparison of the MLCS2K2 fiducial curves (solid), with the Leibundgut template
(dashed) used by \citet{rie99}. The difference between the Leibundgut template and our fiducial 
curve is 1.8 days at about 0.66 mag on the rising portion, which is important in explaining 
the approximately $2$-day difference between our rise time and that of \citet{rie99}. Also, 
application of the single-stretch method using a very wide template will `squeeze' the light curve 
and push the maximum to a later date in order to fit the real rise data. The extrapolations shown here 
are described in Section 4.3; this figure displays 7 of our 13 extrapolations to the MLCS2k2 template.
\label{fiducial}}
\end{figure}

Our implementation of the fitting method uses $B$ and $V$ fiducial curves generated by 
MLCS2K2 \citep{MLCS} for a typical, $\Delta =0$ supernova. The MLCS method has the 
advantage of creating a template from a large number of nearby supernovae. However, few of the 
supernovae used to train MLCS2k2 had light curve information earlier than 10 days before 
maximum light. For the region earlier than $-10$ days, we assume a simple expanding fireball
model which implies that the flux from the moment of explosion increases as the square of the time 
\citep{gold01,rie99}. We then create a set of possible early-time light curves that all join with the 
MLCS2k2 fiducial at $-10$ days (see Figure~\ref{fiducial}). This set ranges in rise time from 14 to 
20 days in half-day intervals. The first derivative of the joined curves is not required to be 
continuous. In Section 4.3, we describe how we use the SDSS-II supernovae to determine which of these template curves is the best match to the observations.

Historically, there are a number of ways to characterize the light curve shapes of SNe~Ia. For 
example, the MLCS2k2 $B$-band fiducial curve falls 1.1 mag in 15 days after maximum, 
corresponding to a \dm15 =1.1 mag \citep{ham96}. Equivalently, the fall time, $t_f$, is defined as 
the number of days it takes a light curve to fade 1.1 mag after maximum brightness. 
The 2-stretch method estimates fall stretch, $s_f$, and the two parameters are simply related by 
$t_f = A\times s_f$, where $A$ is a constant that describes how long it takes the fiducial to fall 1.1 
mag. The relation between \dm15 and $t_f$ depends on the precise shape of the fiducial curve, 
but is approximately quadratic for MLCS2k2: 
$$t_f = 14.711 - 9.631(\Delta m_{15}(B) - 1.1) + 9.391(\Delta m_{15}(B) - 1.1)^2$$
%$$t_f = 36.668 - 30.291\;\Delta m_{15}(B) +9.391[\Delta m_{15}(B)]^2$$ 
over $0.7<$\dm15$<1.5$ for the MLCS2k2 fiducial $B$ band used here. We derived this equation by 
phase-stretching the MLCS2k2 fiducial light curve and directly measuring the $t_f$ and \dm15. Note that 
we use the literal definition of \dm15 and this may not be the same value derived by \dm15 fitters 
\citep{pri06} precisely because rise and fall rates are not absolutely correlated.

%The 2-stretch fitter that we have designed is implemented using the $B$ and $V$-bands and
%applied independently. We estimate the quality of each of these fits separately before averaging 
%the stretches derived from the two bands. The $V$-band peaks later than $B$, but the time of the 
%explosion is the same, resulting in a longer rise time in $V$ than in $B$. In principle, the stretch 
%relative to the fiducial should be the same for the two bands and averaging the two stretches 
%should reduce noise in the rise time estimate.

In order to estimate the rise and fall times of each supernova, an IDL\footnote{Interactive Data 
Language, http://www.ittvis.com/idl/idl7.asp} routine was developed that modifies the fiducial curve, 
$F(\tau)$, to fit the data at rest-frame times, $t$, by performing a $\chi^2$ minimization. 
Mathematically, the 2-stretch function $S$ is defined as
$$ 
  S(t) = \left\{ \begin{array}{rl}
    f_0\; F((t - t_0)/s_r) \;\;\; {\rm if}\; t\leq t_0 \\
    f_0\; F((t - t_0)/s_f) \;\;\; {\rm if}\; t>t_0
 \end{array} \right.
  $$,
where $s_r$ is the rise stretch, $s_f$ is the fall stretch, $t_0$ is the time of maximum light, and $f_0$ 
is the peak flux. The fiducial curve is prepared so peak flux occurs at $\tau =0$ and the peak has a 
value of unity. The function $S$ is an approximation to the true observed light curve and is then 
interpolated at each observation time. The minimization is performed on data out to 25 days past 
maximum light. %For the SDSS-II supernovae the caluclation is done in flux units, but for classic, 
%local supernovae, the data and errors were recorded as magnitudes and we performed the fit in 
%magnitude units.

We applied the 2-stretch fitter to the SNe~Ia studied in \citet{rie99} and a subsample from \citet{hic09}. In 
all, we fit 41 nearby SNe with the 2-stretch algorithm; this sample includes the 8 SNe~Ia analyzed by 
\citet{str07}, 4 more SNe~Ia analyzed by \citet{rie99} that are not in the \citet{str07} set, and 29 SNe~Ia 
from CfA \citep{hic09} that had pre-maximum data. A number of SNe are common to all three data sets, 
and in these cases we use the data with the highest number of pre-maximum observations. SN~1996bo 
from \citet{rie99} is poorly constrained on the rise portion of the light curve, and was not included in our 
analysis. Our $\chi^2$ minimizing function was unable to find a suitable minimum for SN 1997bq and 
SN 1998ef from \citet{rie99}. Lastly, SN 2005hk was not used from the CfA set because it is a peculiar 
SN~Ia, and has been omitted from the SDSS-II analysis as well.

\section{Method}
This section will explain issues not fundamental to the 2-stretch fitter itself, but are important issues 
post-fit in analyzing the data. First, we discuss our error calculation method, and then we explain how 
we use the errors to select our subsample of the `best' SN Ia light curves. Next, we describe our method 
for template selection, ultimately resulting in the selection of a 16.5-day rise-time MLCS2k2 template 
which is used for all 2 stretch fits in this paper. Lastly, we discuss how we combine the $B$- and $V$-
band fits for each SN Ia.

\subsection{Method of Estimating Errors}

The uncertainty in rise and fall measurements arises from both photometric errors and the 
distribution of observations across the light curve. To estimate the rise/fall error, we use a Monte 
Carlo method similar to that of \citet{cont00}. At each observation in a light curve, we employ a  
Gaussian distribution centered on the observed flux and with a standard deviation equal to the flux 
error of that point. We generate simulated photometry at each observed time  and then perform the 
2-stretch $\chi^2$ minimization on the simulated light curve. The randomization is performed 100 
times, and the standard deviation of the set is used as the error for the rise/fall times. There is no 
distinguishable difference in error values if this is performed 100 times or 1000 times. The rise/fall- 
time uncertainties reported by this method are used as a measure of the quality of the fit of each SN. 
%This selection of good SDSS-II light curves using these errors will be described in more detail in \S 
%3.3.

For the low-redshift supernovae, we found that the photometric uncertainties reported in the 
published light curves had a large range even though the supernova apparent peak brightnesses 
were similar. We assume this is due to some authors including systematic errors (photometric 
calibration uncertainties) in some analyses while other studies list only statistical photometric 
uncertainties. We add 0.01 in quadrature to all error values to take into account uncertainty in the 
model. 

We notice a bias toward larger errors on fast-declining supernovae. The reason for this is twofold: 
these supernovae are generally fainter for a given redshift and therefore have larger photometric 
errors, and the rest-frame cadences of observations of these supernovae are longer. By excluding 
supernovae with large rise/fall error estimates, we have reduced the number of very fast decliners in 
our sample relative to supernovae with normal and slow-declining light curves. By a similar 
argument, we expect that the number of broad light curves is over represented in this sample 
when compared to the number of fast-declining events.

\subsection{Selecting Well-Sampled Light Curves}

To determine how the rise and fall times of individual SNe~Ia are related, we had to choose 
supernovae with well-defined light curves before and after maximum light. Sorting supernovae 
based on the number of points within a certain time range does not take into account the variation 
in photometric quality at various redshifts. We decided to select light curves based on the rise- and 
fall-time errors calculated by our Monte Carlo method. 

Since one of our goals was to look for the double-peaked RMF distribution found by \citet{str07}
which has a separation of about 3~days, we decided to use a 2.0-day error cut on the rise time 
and the fall time. We applied these cuts to both the $B$ and $V$ bands before averaging the 
stretches from each band. These criteria produce a sample of 105 supernovae, out of the 391 SDSS-II 
SNe. The redshifts for these events range between 0.037 and 0.230; the average redshift is 0.14. 

\begin{figure}[h!]
\begin{center}
\includegraphics*[scale=0.6]{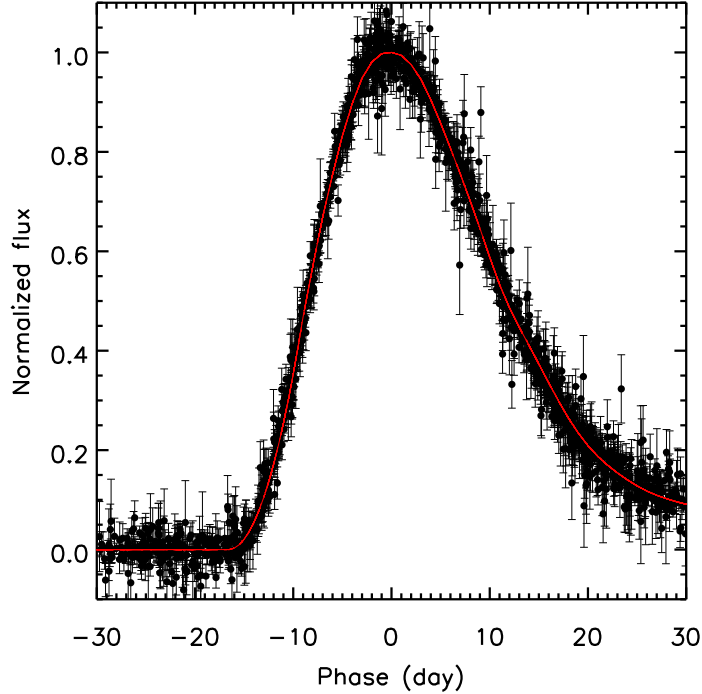}
\end{center}
\caption{105 SNe used in this study, corrected by both the rise and fall stretches,
along with the flux and time of maximum found by the fitter using a 16.5-day rise-time template. The rise 
time of the best template is not the average rise time of the data, but a measure of the best relative 
shape between the extrapolated early rise portion and the MLCS2k2 template. This represents $B$
band only. This normalized light curve has a $\chi^2/dof$ of $211/247$, a probability of fit of 0.988. 
This fit is much better than the single-stretch version, which has a probability of fit of 0.242. 
\label{corrected}}
\end{figure}

Just as in the single-stretch method, it is possible to ``correct" the observed light curve to best 
match the fiducial curve by dividing the time axis by the estimated stretch parameter. This 
normalizes the light curve to match the template curve that was used in the fitting process. The 
only difference for the 2-stretch method is that the pre-maximum and post-maximum time axes 
have different stretches applied.  Figure \ref{corrected} shows the 2-stretch-corrected light curves 
for the 105 SDSS-II light curves well sampled in both rise and fall.

\subsection{Selecting the Fiducial Curve}

The MLCS2k2 fiducial light curves were poorly defined for early phases, so we constructed a set 
of extrapolated curves with a range of explosion dates that meet the MLCS2k2 curve at $-10$ 
days. Recall that in the simple fireball model, the extrapolations are quadratic functions with zero 
flux  at the day of explosion. We created extrapolations at half-day intervals starting with explosion 
at 20 days before $B$ maximum and ending with explosions at 14~days before peak (see Figure~
\ref{fiducial}). 
%We use the 105 SDSS-II light curves that pass our error cuts to choose the best fit 
%extrapolation and we have used this as the fiducial curve for the previous sections in the paper. 

\begin{figure}[h!]
\begin{center}
\includegraphics*[scale=0.6]{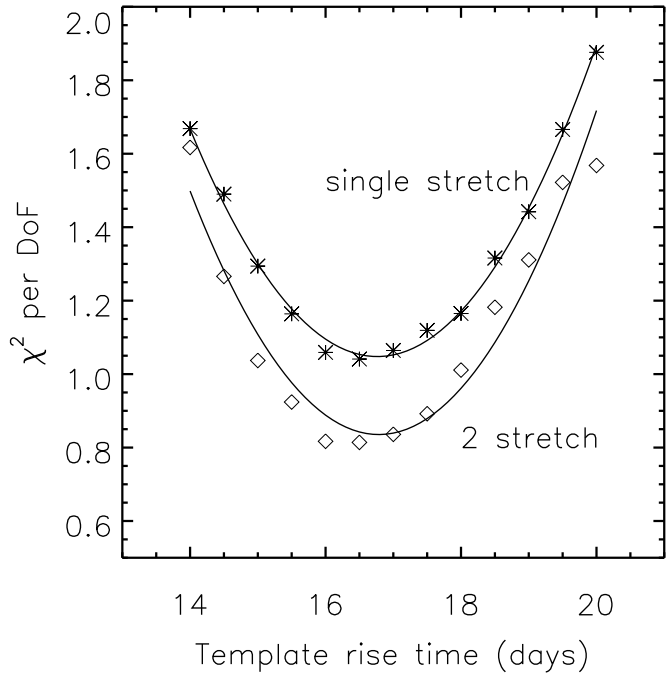}
\end{center}
\caption{Plot of the $\chi^2$ distribution for the fiducial curve rise times, consisting of 105 
high-quality SDSS-II light curves. The light curves were normalized to the fiducial curve using the values 
determined by our 2-stretch fitter, then the reduced $\chi^2$ value of the region between $-20$ and 
$-10$ days was calculated. Plotting the reduced $\chi^2$ for the entire light curve fitting region 
produces a similar result in the comparison of single stretch vs. 2-stretch, but is slightly noisier. 
These data represent the $B$ band only. \label{chi2}}
\end{figure}

To determine the best extrapolation, we fit each of the 105 light curves with the 2-stretch algorithm 
using each of the 13 extrapolated rise curves as a fiducial. We then divide the pre- and post-maximum 
portions of each light curve by the derived rise and fall stretches, and combine all 105 light curves into a 
single normalized light curve for each candidate fiducial curve. We calculate a $\chi^2$ value on the 
data points contained within the region of phase from $-20$ to $-10$ days. We found that the smallest 
reduced $\chi^2$ value and largest corresponding probability of fit were attained with the fiducial curve 
with a 16.5-day rise time in the $B$ band (see Figure \ref{chi2}). This template was best in $V$ band as 
well. This MLCS2k2 16.5-day template is the template that was used for all analyses in this paper. We 
emphasize that this number is not our estimate for the typical rise time of a type Ia SN, but a 
determination of the shape of the early light curve that best matches the data, relative to the shape of the 
actual MLCS2k2 template data later than $-10$ days. We also applied a single-stretch fit and found that 
the probability of fit was always significantly smaller than for the 2-stretch fit, showing that the 2-stretch 
fitting method is a better model for SN Ia light curves (see Figure \ref{chi2}). The $\chi^2$ value goes 
from 1515 (single stretch) to 1140 (2-stretch) with the addition of another model parameter, with 1050 
and 951 degrees of freedom (dof), respectively. At the minimum $\chi^2$, the equivalent probability of fit for 
the 2-stretch fitter is 0.988 while for the single-stretch fitter it is 0.242. % than the assumption of a single 
%stretch parameter for rise and fall time. 

Lastly, we used our error calculations to determine the optimum day after maximum to cut off the fitting 
algorithm. The fraction of  the 391 supernovae that pass our error cuts is a good estimator of the 
overall quality of the fits. We fit all of the light curves with a range of cutoff ages from $+15$ days to 
$+35$ days and found the best cutoff to be around $+25$ days. Longer age cutoffs can bias the fall time 
with noisier data that may not be as well represented by the template; we found that the best 
combination of small errors and negligible bias occurred at the cutoff age of $+25$ days. This age cutoff 
was used in all analyses reported in this paper. 

\subsection{Combining the $B$- and $V$-Band Stretches}

Following \citet{str07}, the stretches from the $B$ and $V$ bands are combined to produce a
single rise and a single fall-time estimate for each supernova. The stretches in the two bands ($s_r(B)$ 
and $s_r(V)$, for example), found using the 16.5-day MLCS2k2 template to find the stretches, are combined using their weighted average, with our derived errors as the weights (in the form $1/ \sigma^2$). This has the advantage of reducing noise in the rise/fall-time measurements. Figure 
\ref{alldata} shows that the SDSS-II $B$-band light curves, on average, decline more slowly than the 
MLCS2k2 fiducial. This could be due to selection bias that comes from the tendency of a magnitude-limited search to find brighter than average events that also tend to have slower-declining light curves. 
However, the average SDSS-II $V$-band light curve is a good match to the MLCS2k2 $V$-band 
fiducial. This mismatch between the average $B$ and $V$ curves and the fiducial is puzzling. We have 
chosen to ignore this difference and simply renormalize the fiducial stretches to best match the SDSS-II 
average light curves in both bands.

We found that the MLCS2K2 templates in $B$ and $V$ were different between their average rise and 
fall stretches when compared to the SDSS-II data. The $V$-band stretches on the fall portion were on 
average smaller than the $B$ band, while on the rise portion the $V$-band stretches were larger than 
$B$ band. In order to renormalize our template, we used the average stretches from the set of 105 
SDSS-II SNe. We divide all $B$- and $V$-band stretches of the actual data by the average $B$-rise, $B
$-fall, $V$-rise, and $V$-fall stretches. This ensures that the templates have the same rise and fall 
stretches in both $B$ and $V$, so that we can average the stretches together with confidence that they 
are centered about the same mean value. This method of modifying the stretches of the data after fitting 
relies on the fact that starting our $\chi^2$ minimizer at a different initial condition will still provide the 
same final result. The other choice is to modify the template before fitting, in which case we would be 
multiplying the template by these average stretches from our data set. As long as the $\chi^2$ minimizer 
finds the same minimum, these two methods are interchangeable. Testing indicated this to be the case.

\begin{figure}[h!]
\begin{tabular}{cc}
\includegraphics*[scale=0.65]{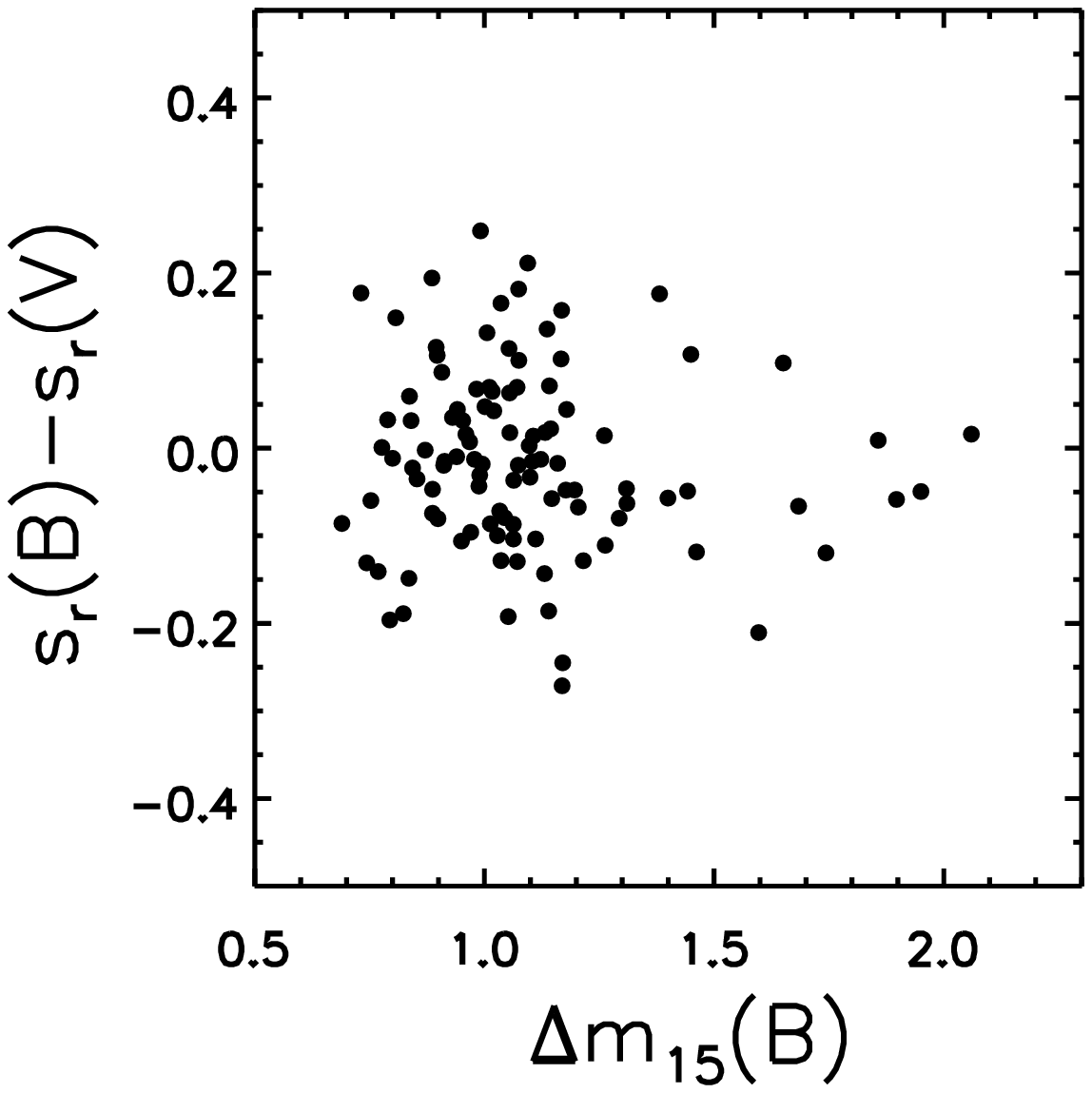} & \includegraphics*[scale=0.65]{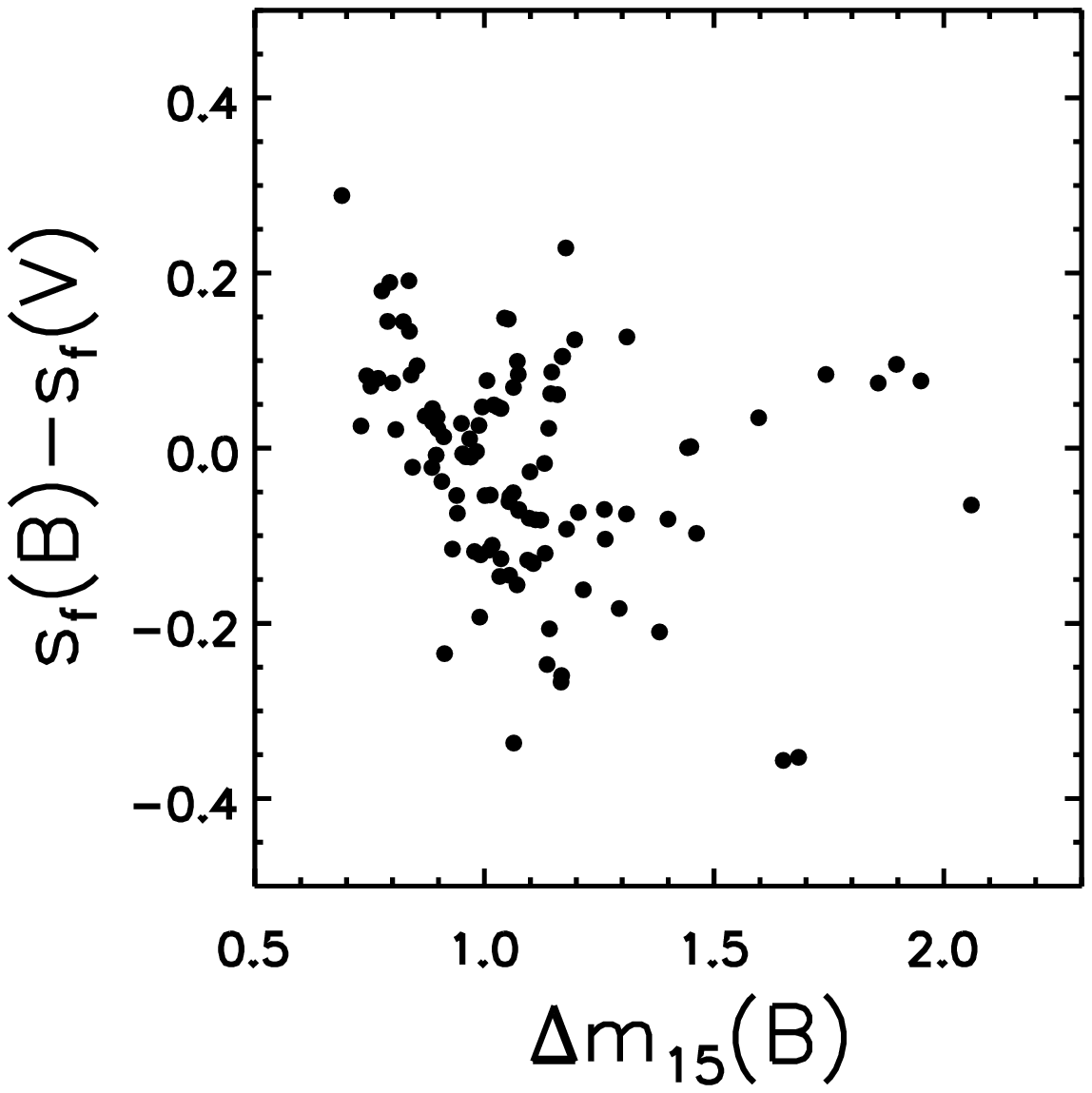}
\end{tabular}
\caption{{\it Left:} the $B$-band rise stretch minus $V$-band rise stretch for 105 SDSS-II type Ia SNe. 
The selection criterion for these 105 is described in Section 4.2. The $B$- and $V$-band rise stretches show
no clear variation across the sample, indicating that the rise stretch is consistent between the two bands.
{\it Right:} the difference between the $B$-band fall stretch and the $V$-band fall stretch for the same 
105 SNe. Here, the data show evidence for variation with \dm15. Specifically, we find that 
slow-declining events have a larger than expected $B$-stretch than $V$-stretch when compared to 
events with \dm15$ > 1.1$. 
\label{strdiff}}
\end{figure}

To clarify the relation between $B$ and $V$ light curves, the difference between the rise stretches and 
fall stretches after renormalizing the templates for the two bands is plotted in Figure \ref{strdiff}. We 
find that on the rise, the $B$- and $V$-band stretch difference is constant with decline rate, suggesting 
that the shape of the rise portion of the fiducial curve is consistent between the two wavelengths. But the 
plot of $s_f(B)-s_f(V)$ shows a kink near \dm15=0.9, indicating that the fading portions of the $B$ and 
$V$ light curves require two different stretch parameters to match the data. Specifically, very slow 
declining events tend to fade more slowly in the blue than in the visual band when compared to normal 
decliners. This effect should make very slow decliners more blue after maximum than their faster-fading 
cousins.

Clearly, the $B$-band fall stretch is not strictly correlated with the $V$-band fall stretch
and this may indicate a physical difference between the slowest fading events and more
normal-declining supernovae. However, the effect is small enough that we do average the
derived stretches for this analysis.

\section{Analysis}

\subsection{Rise versus Fall Times in the SDSS-II}

\begin{figure}[h!]
\begin{tabular}{cc}
\includegraphics*[scale=0.65]{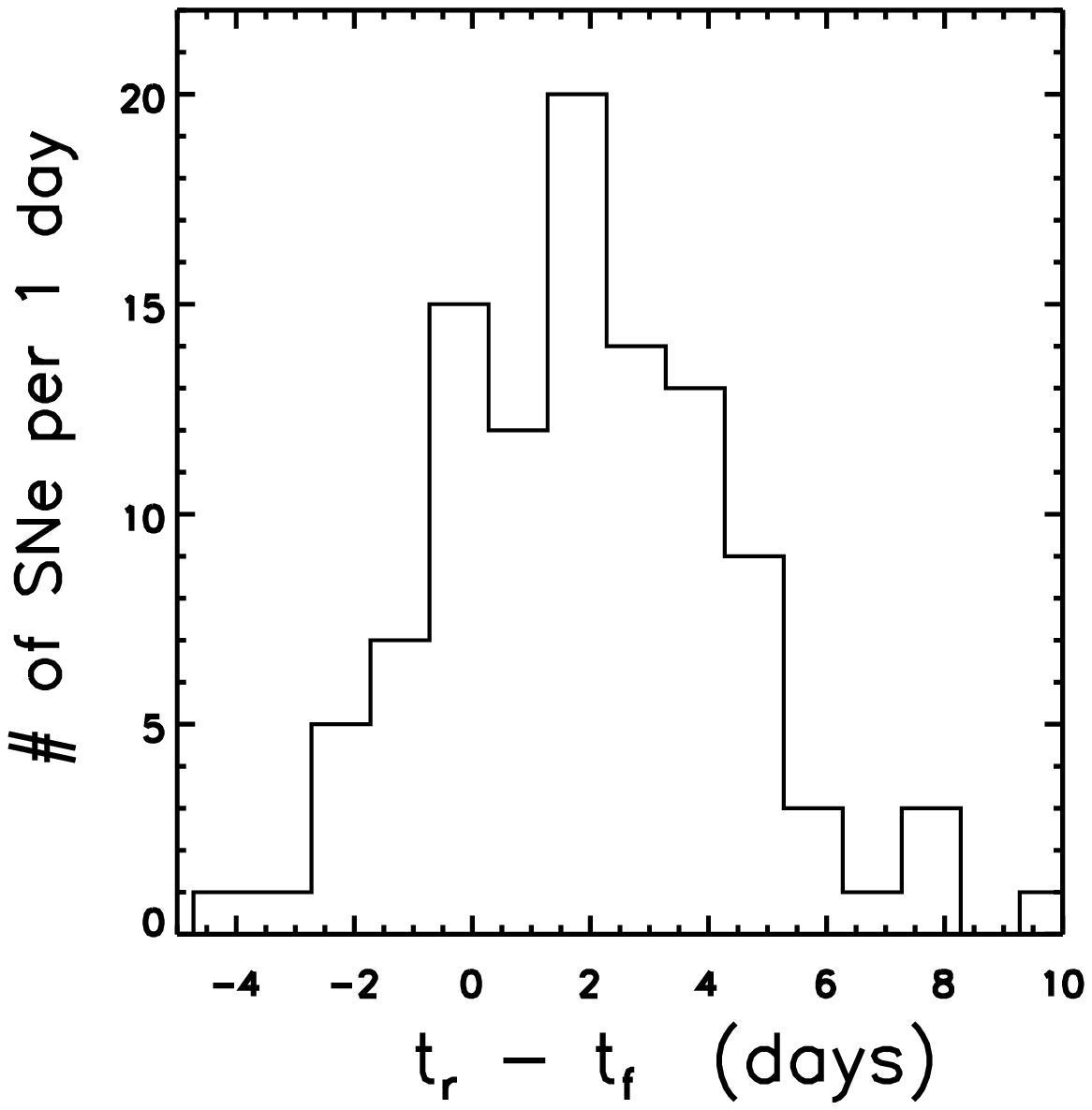} & \includegraphics*[scale=0.65]{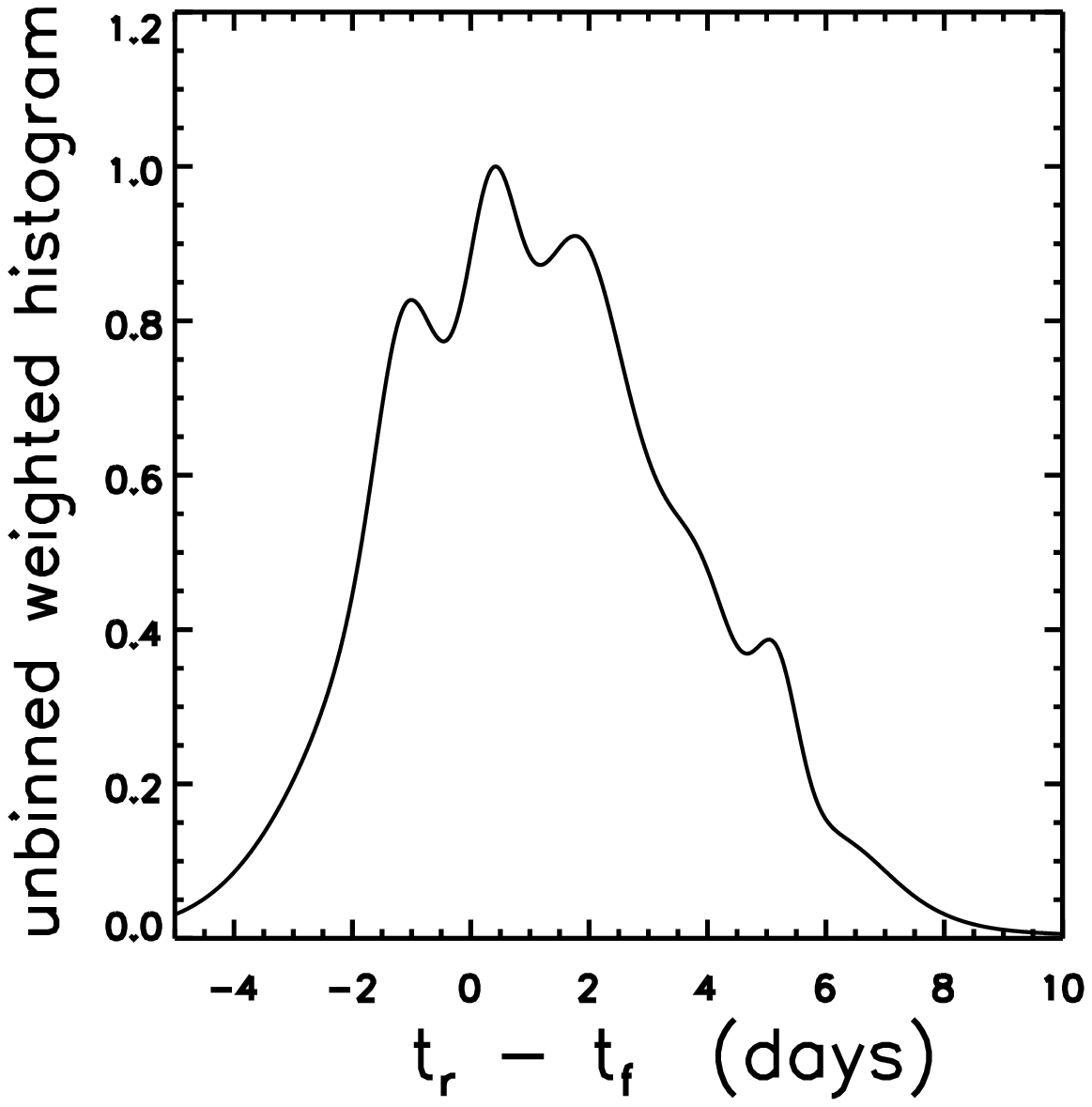}\\
\end{tabular}
\caption{{\it Left:} standard histogram of the rest-frame $t_r-t_f$ distribution for 105 SDSS-II SNe. 
This figure is suggestive of only a single class of light curves, at least in terms of their $t_r-t_f$ 
values.  {\it Right:} the ideogram of $t_r-t_f$ values, where each supernova is treated as a Gaussian 
and then these Gaussians are added together. This is also representative of a single distribution in 
$t_r-t_f$, and agrees very well with the standard histogram. The ideogram is described in detail in 
Section 5.1. 
\label{hist}}
\end{figure}

\begin{figure}[h!]
\begin{tabular}{cc}
\includegraphics*[scale=0.47]{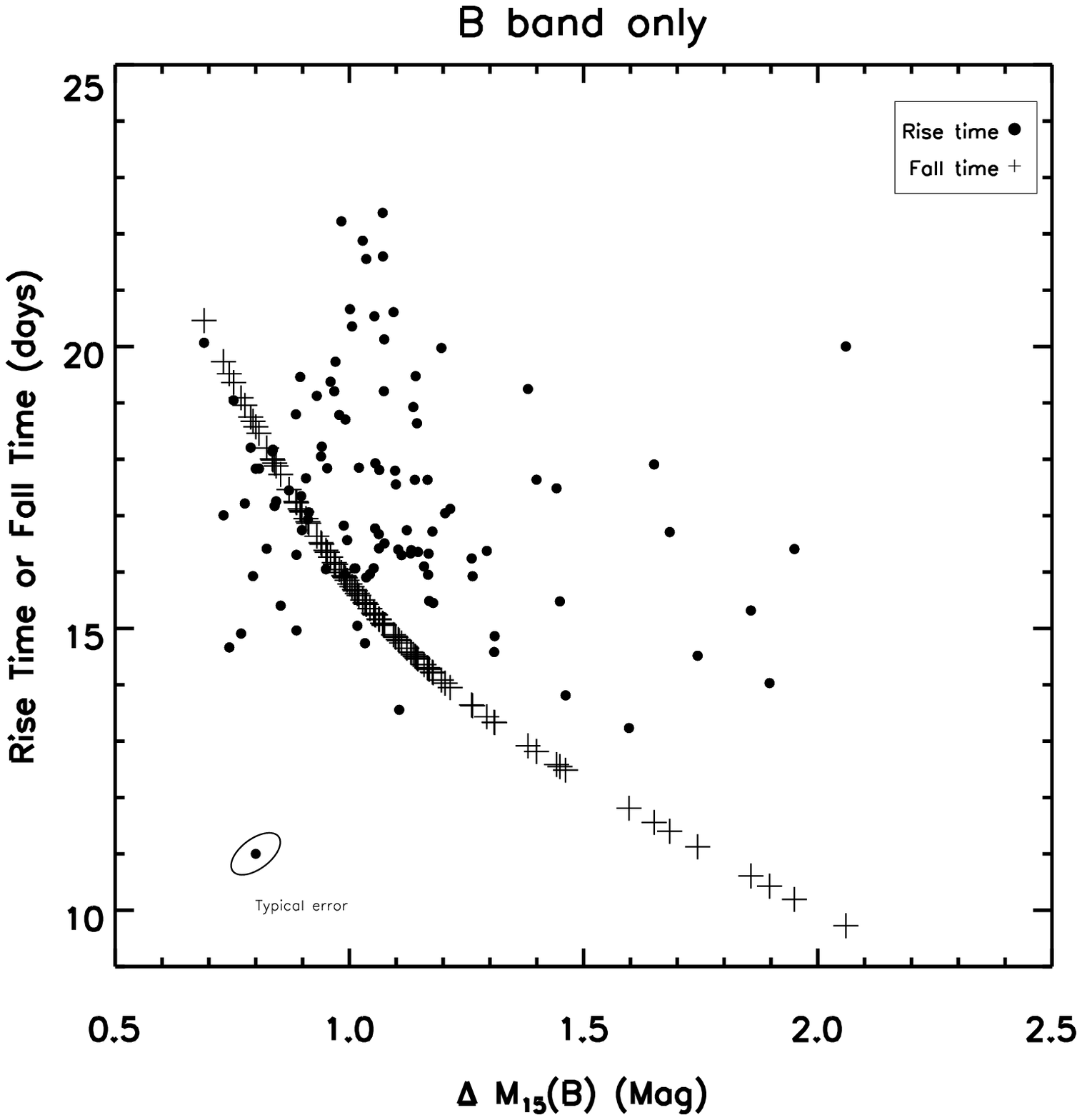} & \includegraphics*[scale=0.47]{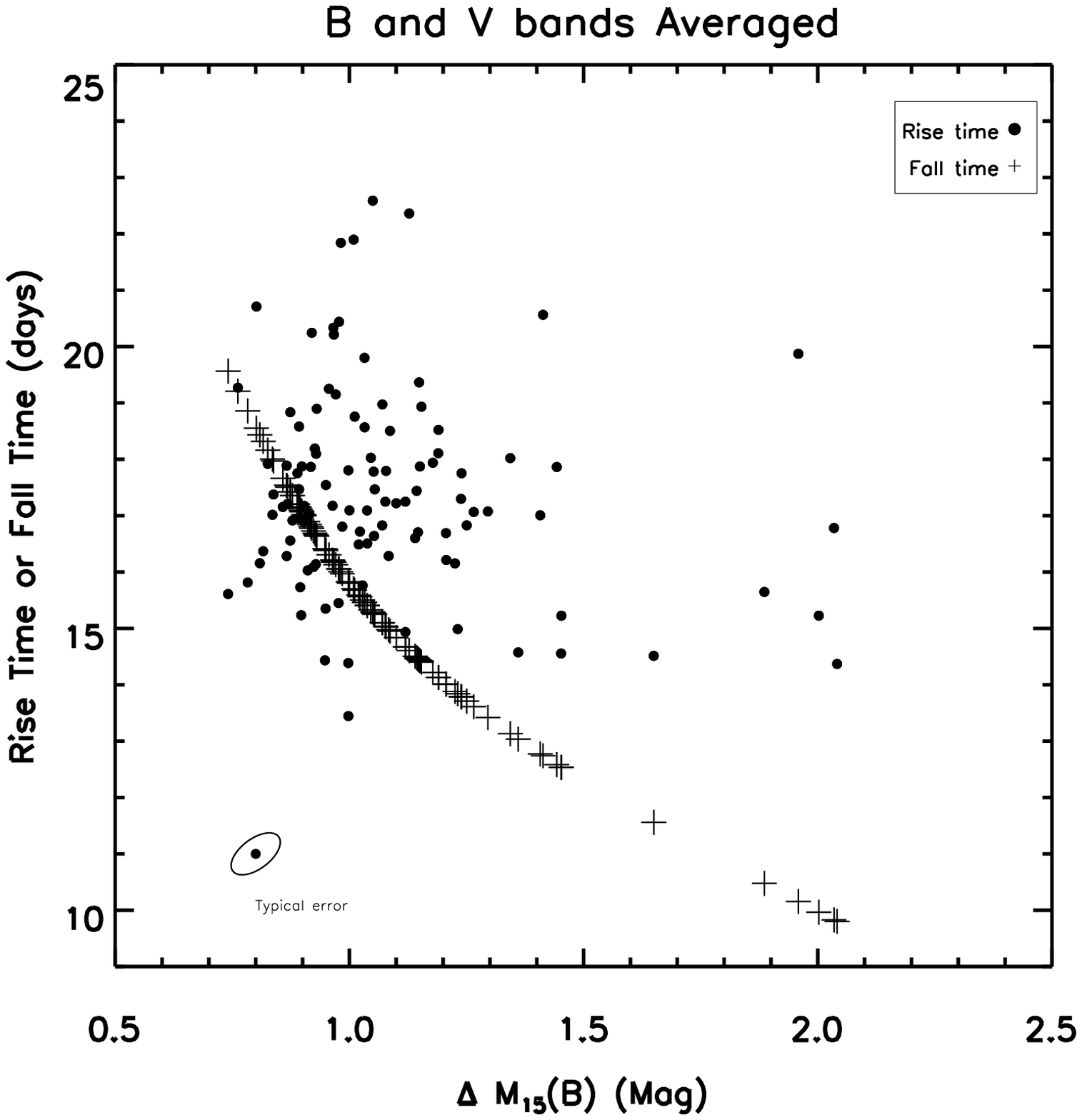}
\end{tabular}
\caption{{ \it Left:} the rise (dots) and fall times (crosses) vs. \dm15\ for the 105 SDSS-II supernovae. 
Fall times are directly related to \dm15 and the crosses show the relation. These data are for $B$ band 
only, and are shown to reinforce that averaging the stretches introduces no bias to the data. The typical 
error ellipse for the rise time or fall time is shown in the lower right. The ellipse results from the 
covariance of the rise and fall estimates, resulting from the error in measuring the time of maximum 
brightness. {\it Right:} here, the $B$- and $V$-band stretch estimates have been combined.  Of special 
interest in this figure is the tendency for the slowest declining SNe to be among the fastest risers, 
however, there are very few slow-declining slow-rising SNe found in the sample. Also of note is a 
minimum rise time of about 13.5 days.
\label{sdss}}    
\end{figure}

Using our sample of 105 high-quality SDSS-II supernovae, with the 16.5-day MLCS2k2 template for the 
fits, we obtain an RMF distribution that is best described by a single Gaussian distribution.  This 
distribution can be seen in Figure \ref{hist}, which shows a standard histogram of the RMF distribution 
along with an ideogram of the distribution \footnote{Particle Data Group, http://pdg.lbl.gov/2007/reviews/
textrpp.pdf}. These histograms, along with the actual rise and fall distributions shown in Figure 
\ref{sdss}, are suggestive of a broad, single-peaked distribution in RMF for the SDSS-II supernovae. The 
width of the RMF histogram is wide compared to the errors in the measured rise and fall times. This 
implies that the rise time is not strongly correlated with the fall.

The ideogram in the right panel of Figure \ref{hist} is a visual device only and is used to display a 
histogram of the data weighted by errors in $t_r - t_f$. It is not used to numerically evaluate any 
properties of the data. The distribution of a standard histogram, one where the data values are 
separated into rigid bins and then simply counted, can be very dependent upon the location and 
width of the bins.  The ideogram, however, treats each data point as a Gaussian distributed about 
the mean value, so that the appearance of the distribution is only dependent upon the data values 
and their errors. For our ideogram, we weight the area that each RMF value contributes to the 
overall distribution by $1/\sigma$. In other words, instead of normalizing the Gaussian distribution 
to $1$, we normalize to $1/ \sigma$. For each supernova RMF value, this gives us the distribution:
$$ 
D_i(x) = \frac{1}{\sigma_i^2 \sqrt{2 \pi}}e^{-\frac{\left(x-{RMF}_
i\right)^2}{2 \sigma_i^2}}
$$,
where $\sigma_i$ is the error in ${RMF}_i$. The ideogram is then simply given by $I(x) = 
\displaystyle\sum_{i=1}^N D_i(x)$.

\begin{figure}[h!]
\begin{center}
\includegraphics*[scale=0.6]{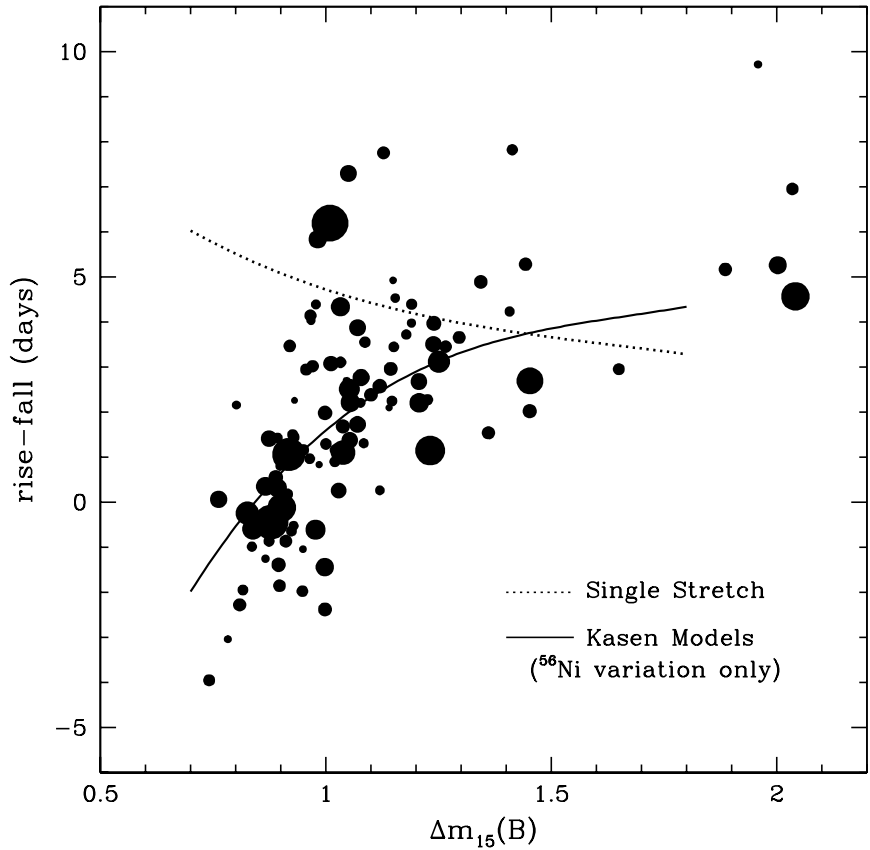}
\end{center}
\caption{Plot of $t_{r} - t_{f}$ against \dm15, showing only the 105 SDSS-II supernovae. This 
figure displays the unexpected result that the slowest declining SNe tend to be the fastest risers. 
The dotted line shows $t_r-t_f$ vs. \dm15 relation if SNe~Ia followed a single-stretch model with a 
rise time of 19.5 days and a fall of 15 days. The solid line shows the predicted $t_r-t_f$ for a series 
of Kasen light curve models that vary total radioactive nickel yield. The size of the points represents 
the error, so that  larger points have smaller error. \label{rise-fall}}
\end{figure}

The $t_r-t_f$ values are not uniformly distributed with fall time (\dm15 ). Figure~\ref{rise-fall} 
displays the RMF versus \dm15 for the 105 supernovae. We see that there are few slowly declining 
light curves (\dm15$<1.0$ mag) with large RMF values. In other words, many of the slowest 
declining SNe are among the fastest risers. This point is reiterated by the actual distribution of rise 
and fall shown in Figure \ref{sdss}.

For comparison with the SDSS-II data, we plot in Figure~\ref{rise-fall} the RMF trajectory of a  
19.5-day rise-time fiducial curve with a single stretch applied to both rise and fall. This is currently 
the `standard model' of a SN~Ia light curve and it does pass through much of the real data; however, 
the standard model cannot match the fast rising or very slow rising light curves in the SDSS-II 
sample.

\subsection{Average Rise Time and Decline-Rate-Corrected Curves}

Using our errors to weight the rise times of the sample, we obtain for the nearby set of SNe~Ia an 
average rise time of $16.82 \pm 0.28$ days (standard error) with a standard deviation of $1.77$ days. 
For the SDSS-II SNe~Ia, we obtain a weighted average rise time of $17.38 \pm 0.17$ days (standard 
error) with a standard deviation of $1.8$ days. These values are in general agreement, especially 
considering that the SDSS-II supernovae and the local sample have different selection
biases.  This result agrees very well with that
of \citet{str07}, who found an average rise time of $17.44 \pm 0.39$ days in the nearby sample.

In order to gain further insight into the SDSS-II light curves, it is beneficial to perform a decline-rate 
correction to the sample of light curves, similar to the procedure described by \citet{str07}. This 
correction is performed by dividing the entire time axis of each SN by the fall stretch found by the 
2-stretch fitter. Based on the ideas of the single-stretch method, that every SN~Ia light curve maintains 
the same relative shape between rise and fall, this approach should normalize the entire light curve to 
the shape of our template. 

We applied a decline-rate correction to the 105 SNe in our data set, by dividing the entire time axis 
by $s_f$ only, and found that the spread of the rise portion of the light curves increased.  For the 
SDSS-II SNe~Ia, the average decline-rate-corrected rise time is $17.32 \pm 0.12$ with a standard 
deviation of $2.89$ days. The increase in the spread of the rise times after applying the decline-rate 
correction is a departure from the expected result implied by a single-stretch method. 

\begin{figure}[h!]
\begin{tabular}{r}
\includegraphics*[scale=0.47]{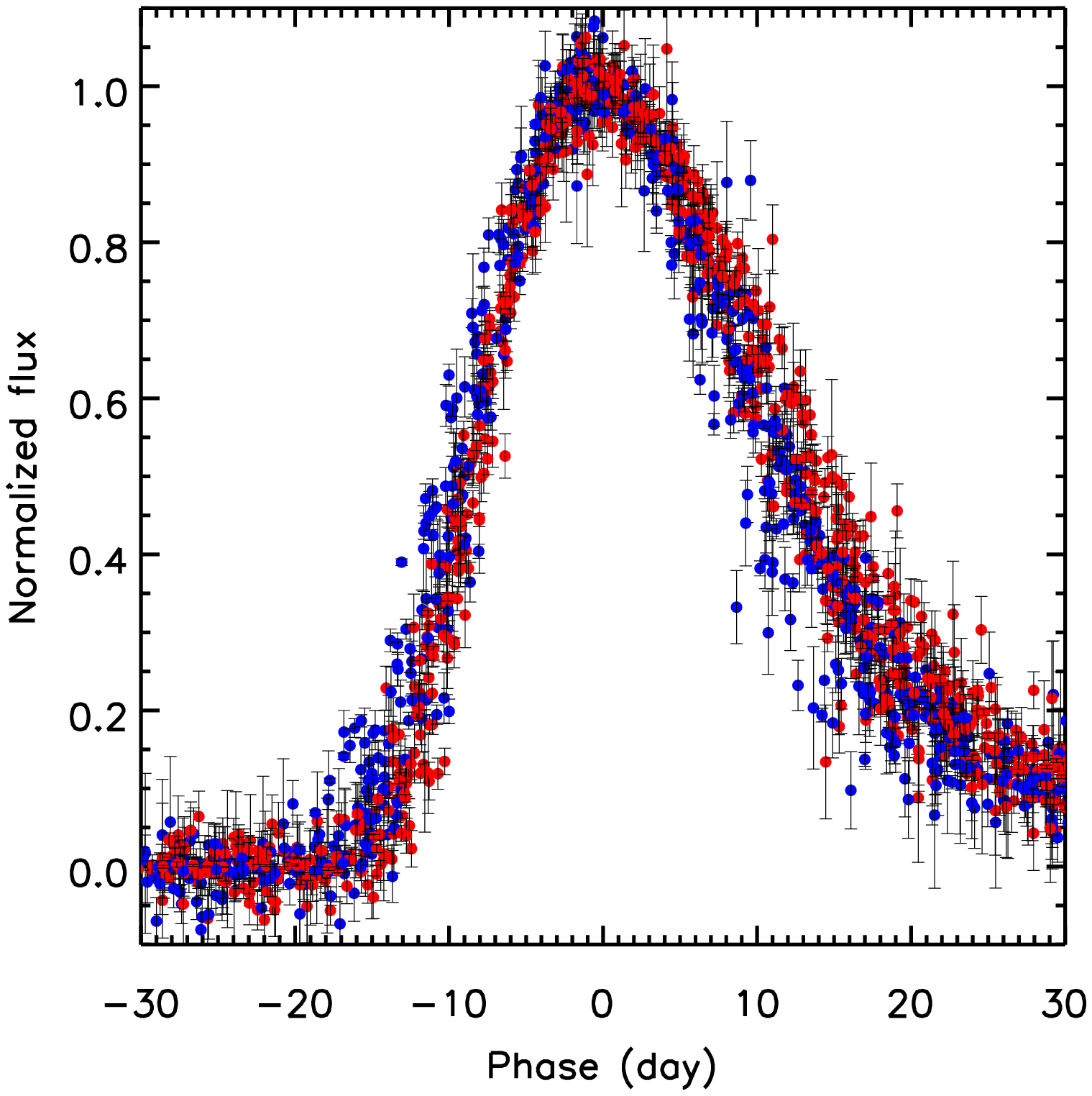}
\includegraphics*[scale=0.47]{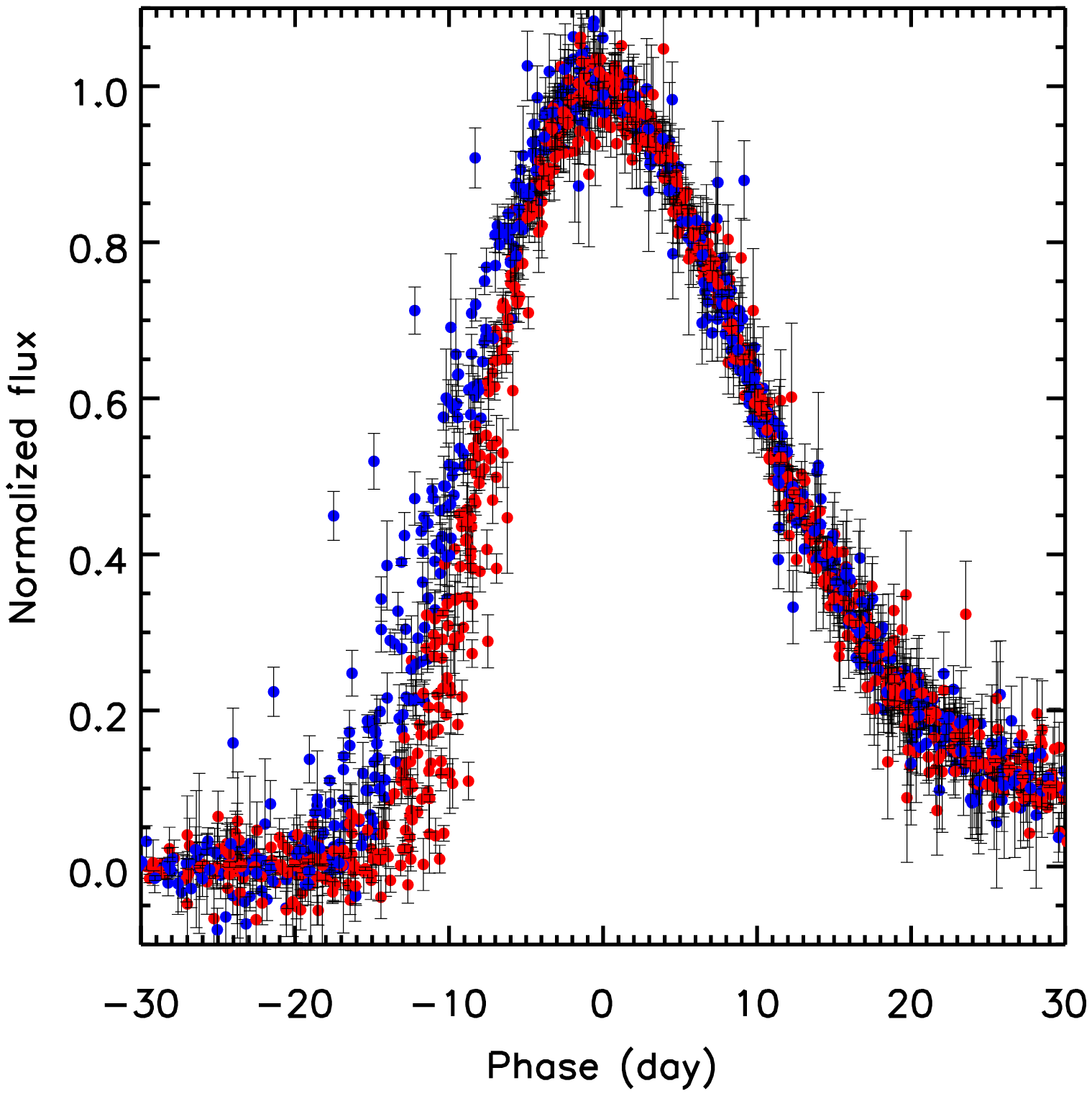}
\end{tabular}
\caption{{\it Left:} the 105 SNe~Ia used in this study, with no stretch corrections applied and $B$-band 
data only. The red points represent SNe~Ia with $t_r-t_f \leq 2.0$ days, while the blue points represent 
SNe~Ia with $t_r-t_f > 2.0$ days. {\it Right:} the same supernovae, with the entire time axis divided by 
the fall stretch as found by the 2 stretch fitter, showing the tendency for the slowest decliners to be the 
fastest risers. This analysis implies that the previously accepted correlation between rise and fall, e.g., a 
single-stretch model, is not supported by the data once the rise and fall times are disconnected in the 
fitting process. \label{deccorr}}
\end{figure}

This increase in the rise-time spread can be seen graphically in Figure~\ref{deccorr}, which shows the 
result of applying the decline-rate correction to the 105 SNe~Ia that pass our error cuts. This plot uses 
$B$-band data only, along with $B$ band only stretches, not the weighted average that was used in 
the RMF analysis. The SNe plotted in red have RMF $\leq 2.0$ days, while the ones plotted in blue 
have  RMF $> 2.0$ days (the color selection for these supernovae is not meant to imply anything 
about the intrinsic colors). The plot on the left of Figure \ref{deccorr} contains the SNe with no stretch 
corrections. Even without stretch corrections, the data show the distinct light curve shape 
differences in SNe~Ia that are also shown graphically in Figure \ref{rise-fall}.   Without stretch correcting, 
we see clearly that most of the slowest declining supernovae are also the fastest rising supernovae; 
this effect becomes more apparent after decline-rate correction.  The slow-declining, fast-rising light 
curves (in red) rise even faster, while the faster declining, slower rising light curves (in blue) rise 
more slowly. The dispersion in rise time increases from the uncorrected mean value, since the 
average rise times agree between the corrected and uncorrected distributions. This is more 
evidence that current SN~Ia light curve fitting models are not able to acceptably characterize the full 
range of SN~Ia light curve shapes. We have arbitrarily selected the dividing point at 2 days in $t_r-t_f
$, which is the mean value, and we emphasize that the light curves show a continuum of $t_r-t_f$ 
values displaying minimal evidence for a two-group classification using this parameter.

\subsection{Comparing the 2-Stretch Fitter with Single Stretch}

In order to further examine the result that the 2-stretch fitter is a better representation of SN Ia 
light curve shapes, we selected a subset of 99 SDSS-II SNe~Ia (from the 105 that passed our 2-stretch 
only error cuts) that passed our error cuts using both the 2-stretch fitter and a single-stretch fitter, 
and calculated the $\chi^2$ value of the stretch-corrected light curve. In this analysis, we calculated 
$\chi^2$ over the entire region that was fit. For the 2-stretch fitter, this composite light curve has a $
\chi^2$ value of 1140 with 951 dof. For the single-stretch fitter, this composite 
light curve has a $\chi^2$ value of 1515 with 1050 dof.  Adding another parameter 
to a fitting procedure should be expected to reduce the $\chi^2$ value by approximately 1 (for each 
individual fit, so in this case 99) if that fitting parameter does not add any new useful information to the 
results. In our case, the 2-stretch fitter has a significant impact in reducing the $\chi^2$ value of the 
composite light curve, demonstrating that the 2-stretch fitter is a better model for representing SN~Ia 
light curves than a single-stretch fitter.
% its possible the DoF are actually 1050 (1s) and 951 (2s)

\subsection{Fitting Simulated Light Curves}

As part of the SDSS-II supernova analysis, a software package called SNANA was developed to 
analyze data and to create simulated light curves \citep{kess09b}. SNANA can create a set of 
supernova light curves with the same cadence, redshift distribution, and photometric signal-to-noise 
ratio as the SDSS-II survey, created from a variety of light curve models. In this section, we test the 
MLCS2k2 model and the single-stretch model against the real SDSS-II SN Ia data. 

\begin{figure}[h!]
\begin{tabular}{cc}
\includegraphics*[scale=0.47]{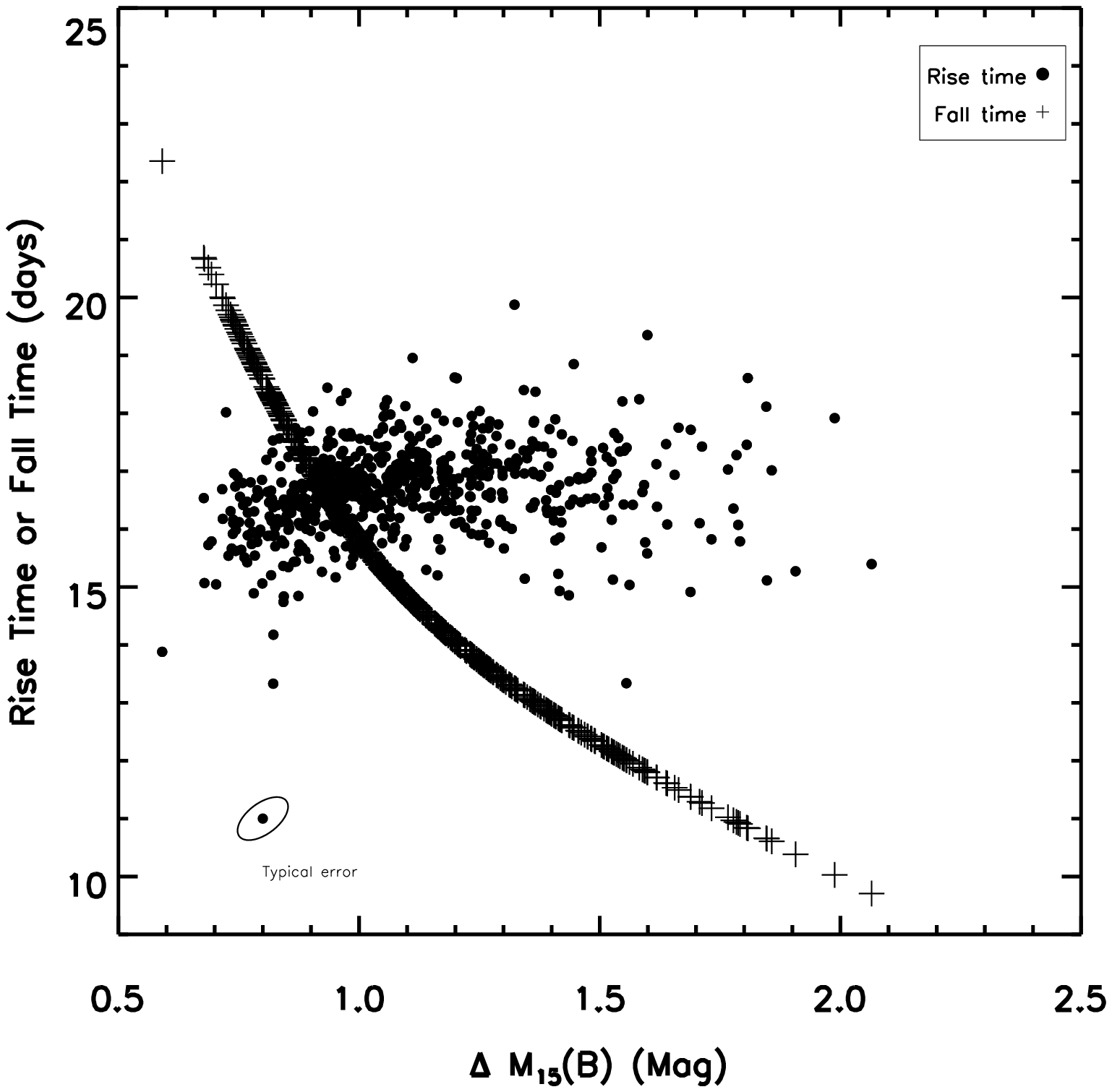}  & \includegraphics*[scale = 0.47]{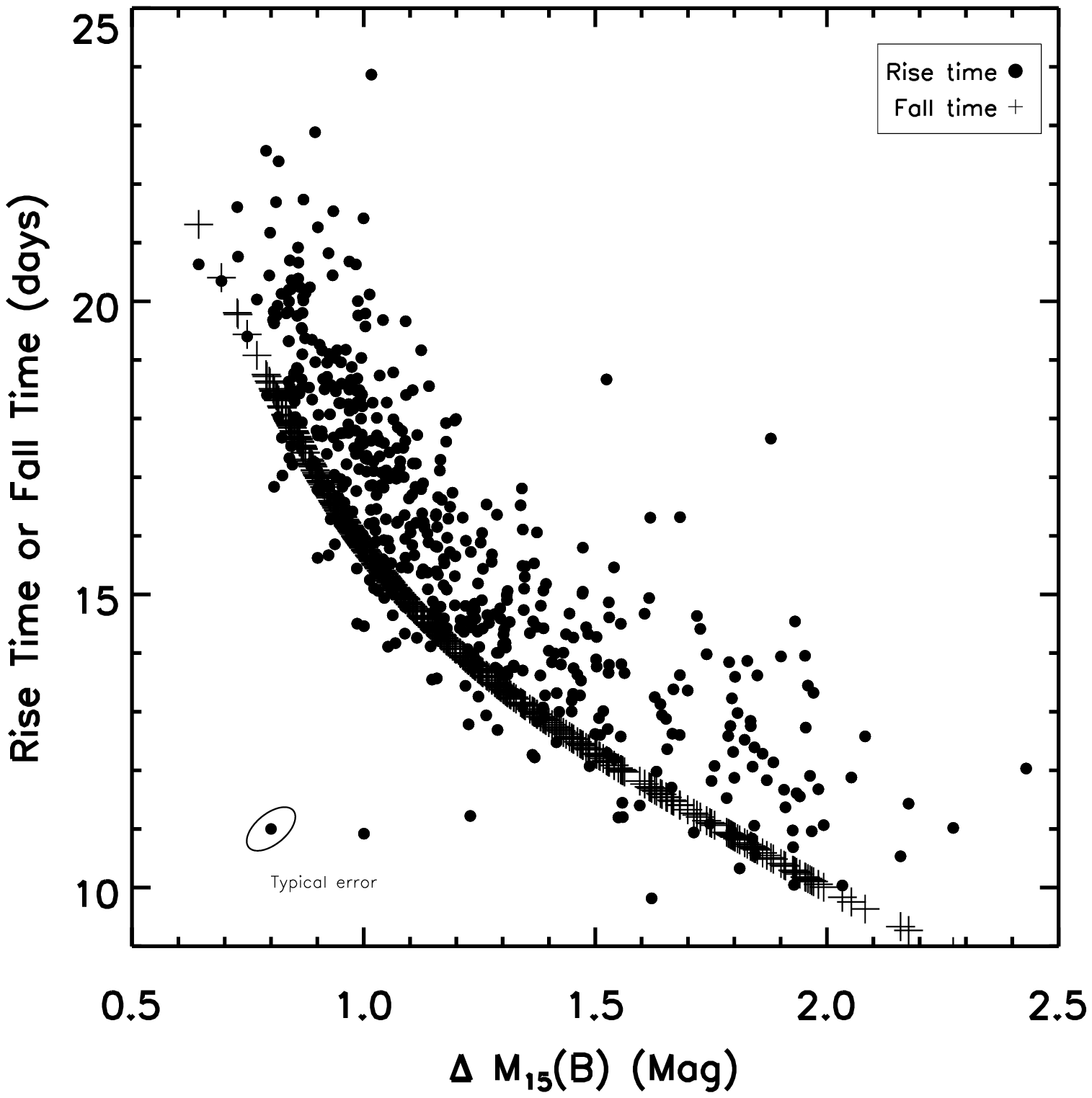}
\end{tabular}
\caption{ {\it Left:} distribution using SNANA simulated curves modeled with the MLCS2K2 17.0-day 
rise-time template. From an initial set of 2694 simulated light curves, 620 pass our error cuts, a very 
similar fraction as for the real data. Note that the MLCS2K2 model does not capture the true spread 
in rise times observed in the SDSS-II data. The rise time stays mostly constant, except in slowly 
declining SNe where it decreases with increasing fall time, an interesting correlation between rise and 
fall that is observed in the SDSS SNe as well. {\it Right:} simulated curves from the same input template 
but for a single-stretch model, and fit with the 2-stretch fitter. The rise time follows the fall time exactly for 
all values of decline rate, and there is no apparent trend for slow-declining light curves to have fast rise 
times in this single-stretch simulation.\label{sim1}}
\end{figure}

We synthesized 2694 simulated light curves using a 17-day rise-time MLCS2K2 template with a 
range of $\Delta$ values and extinction that matches the analysis of SDSS-II SNe~Ia done in 
\citet{kessler08} (see Section 6). These simulated curves have a range of $z$-values from $0.019$ to 
$0.35$ with an average of $0.19$. We applied the same error cuts to these simulated curves as we 
applied to the SDSS-II data; the $z$-values after error cutting range from $0.019$ to $0.24$ with an 
average of $0.12$. This simulated distribution after error cutting is shown in Figure \ref{sim1}. It 
displays characteristics that are present in the SDSS-II SNe as well; it contains many SNe that are 
very slow decliners but are among the fastest risers. However, the simulation is lacking SNe that 
have relatively normal fall times yet are slow risers such as SN~1990N.

We also synthesized 2651 curves using a 17-day rise-time template simulated with a single-stretch 
model. The stretch values are based on the same analysis from \citet{kessler08}. The $z$-values 
range from $0.019$ to $0.35$ with an average of $0.25$. After error cutting, the $z$-values range from 
$0.019$ to $0.34$ with an average of $0.18$. We fit these simulated curves with the 2-stretch fitter, 
but their underlying distribution is based on single stretch. The results can be seen in Figure 
\ref{sim1}. As expected, this single-stretch simulation shows a strong correspondence between rise and 
fall that is not apparent in the SDSS-II data. Upon comparison with Figure \ref{sdss}, there is a 
noticeable difference in the spread of RMF values, with the SDSS-II data having a much greater range. 
Once again this implies a lack of correlation between rise and fall that is unachievable in this single-stretch model of light curve generation. However, the single-stretch simulation does produce some 
normal-declining events that have very slow rise times.

Our simulated curves using an MLCS2k2 model and a single-stretch model imply that the 
range of SN~Ia light curves is not fully matched with current light curve fitters. MLCS2K2 
models appear to be slightly better than single-stretch models as they capture some of the slow-
declining fast-rising SNe~Ia that are evident in the real data. Indeed we expect MLCS2K2 to be better 
simply because it is trained on real SNe. Also note that the MLCS2k2 synthesized curves maintain 
an approximately constant minimum for the rise time of fast-declining SNe at around 13-14 days, 
which appears in the SDSS-II SNe but could be a result of meager numbers of these SNe~Ia. This 
could be an indication of a physical constraint on the physics of SN~Ia explosions, as it seems that 
increasingly fast decliners do not correspond to increasingly fast risers. 

\section{Discussion}

\subsection{Comparison to Previous Rise-Time Studies}
We obtain an average rise time for the SDSS-II SNe~Ia of $17.38 \pm 0.17$ days (standard error of 
the mean); this is significantly different from the result obtained in \citet{rie99}. For the entire nearby 
sample, we obtain an average rise time of $16.82 \pm 0.28$ days (standard error). The standard 
deviation in the rise time is $1.77$ days. This is in general agreement with the value reported for the 
SDSS-II data. The reason that our results differ from the 19.5 days found by \citet{rie99} is the 
difference in shape of the fiducial curves used in the two studies, and the application of the single-stretch fitter in previous methods. \citet{ald00} prove that even slight variations in the declining portion 
of the template can change the measured rise time by 2 or more days using a single-stretch method. 
Figure \ref{fiducial} shows that the ``Leibundgut" template \citep{leib89} used by \citet{rie99} differs 
greatly from the  MLCS2K2 curves in the pre-maximum phase. The Leibundgut template is 0.66 
mag fainter than maximum light at $-10$ days while the MLCS2k2 fiducial curve reaches 0.66 
mag at $-8.2$ days. Such a broad template combined with a single-stretch fit tends to force the 
rise to be very slow and to push the estimated time of maximum later than the true peak brightness (see 
Figure 7 in \citet{rie99}). The narrower MLCS2k2 fiducial combined with the 2-stretch fitter shows that the 
rise is faster than that found by several previous studies and that there is a wide range of observed rise 
times for a given fall time.

We created extrapolations to the Leibundgut template in the exact same manner as for the MLCS2k2 
template, ranging in rise time from 14 to 20 days in half-day intervals. Using the template determination 
method described in Section 4.3, we found the best extrapolated Leibundgut template had a rise time of 19.5 
days (note that this utilizes the 2-stretch fitter). However, after our error cutting, we obtain an average 
rise time of $17.18 \pm 0.19$ days using this template. Even though the extrapolated Leibundgut 
template has a longer rise time than the MLCS2k2 template, the stretches obtained are smaller so that 
the average is consistent. Indeed this should be expected with the 2 stretch fitter; the template with the 
best relationship between early rise and later rise is selected, but the overall rise time remains constant 
despite the difference in the fiducial relationship between rise and fall.

To further demonstrate this point, we also performed a rise-time extrapolation using the same method as 
\citet{rie99}, as well as using the Leibundgut template. We fit all 391 SDSS-II SNe Ia with a single-stretch 
fit using only data from -10 days to 25 days, and we cut SNe with stretch errors greater than 2.0 days.  
After stretch correcting the full light curve, we used the data less than -10 days (which was not used in 
the fitting process) to fit a parabolic extrapolation to explosion. Using this method, we obtain an 
extrapolation of 19.64 days for the SDSS-II SNe Ia. This agrees quite well with both the results of 
\citet{rie99} and our own template selection using Leibundgut template extrapolations and the 2 stretch 
fitter. 

We have shown in this section that the difference between our rise time and that of previous studies 
revolves around the flexibility of the 2 stretch fitter regarding the template used. A single-stretch fitter that 
is used to determine the best extrapolation to explosion is much too reliant on the relationship between 
rise and fall in the template. The 2 stretch fitter is more flexible in the sense that the input template can 
have any relationship between rise and fall and the average rise time will be consistent with other 
templates. Performing a single-stretch fit and extrapolating to explosion gives the same result as our 
template determination method from Section 4.3. However, in order to find the average rise time, the addition 
of another parameter to independently estimate the rise stretch is required, because the rise and fall are 
not strictly correlated. 

\subsection{Shape of the Early Rise}

As with previous studies, we have assumed that the optical flux rises as $t^2$ soon after explosion. 
\citet{arnett82} showed that adiabatic losses should nearly balance heating by radioactive decay and 
keep the effective temperature relatively constant. So the luminosity goes as the radius squared and 
therefore the time squared for a constant expansion rate. \citet{con06} directly tested this model by fitting 
light curves with the temporal power-law index as a free parameter. They found the temporal index $n$ 
that best matched the low- and high-redshift supernovae was 1.8$\pm0.2$, consistent with the Arnett 
calculation. 

To estimate the shape of the early rise, we fit the 105 SDSS-II light curves with good rise and 
fall data with the 2-stretch fitter using only epochs later than $-10$~days. We then corrected all the light 
curves to the fiducial curve using the rise and fall stretches which provides 103 $B$-band observations 
between $-20$ and $-10$ days before maximum. We then calculate the $\chi^2$ parameter over this 
interval by comparing the data to the function $f=A(t-t_0)^n$, where the time of explosion ($t_0$) and 
power-law index ($n$) are allowed to float. The value of the parameter $A$ is set by the condition that 
the function must meet the MLCS2k2 fiducial curve at $-10$~days. 

We found the best-fit power-law index for the SDSS-II data was $n=1.80^{+0.23}_{-0.18}$. The index is 
strongly correlated with the time of explosion which we found to be $-16.8^{+0.9}_{-0.6}$ days. The 
result is consistent with the Arnett prediction and the \citet{con06} estimate for the early light curve 
shape.

We can also test whether or not the early color of the SDSS-II supernovae follows Arnett's prediction of 
slow and modest temperature changes. For each SDSS-II observation where there is a good rest-frame 
$B$ and $V$ measurement, we construct a color based on the $V-B$ normalized flux difference. This 
essentially compares the supernova color to the color at $B$-band maximum and avoids the problem of 
dealing with magnitudes at low flux levels. 

As a consistency check, for each supernova we take the flux color averaged between $-15$ and $-10$ 
days to see how it varies with redshift. We find that the color is essentially constant as a function of 
redshift which gives us confidence that the SNANA $k$-corrections at early times are self-consistent. We 
also see that the colors for $z>0.3$ are very noisy and we restrict our color analysis to the 332 events at 
redshifts less than 0.3.

\begin{figure}[h!]
\begin{center}
\includegraphics*[scale=0.6]{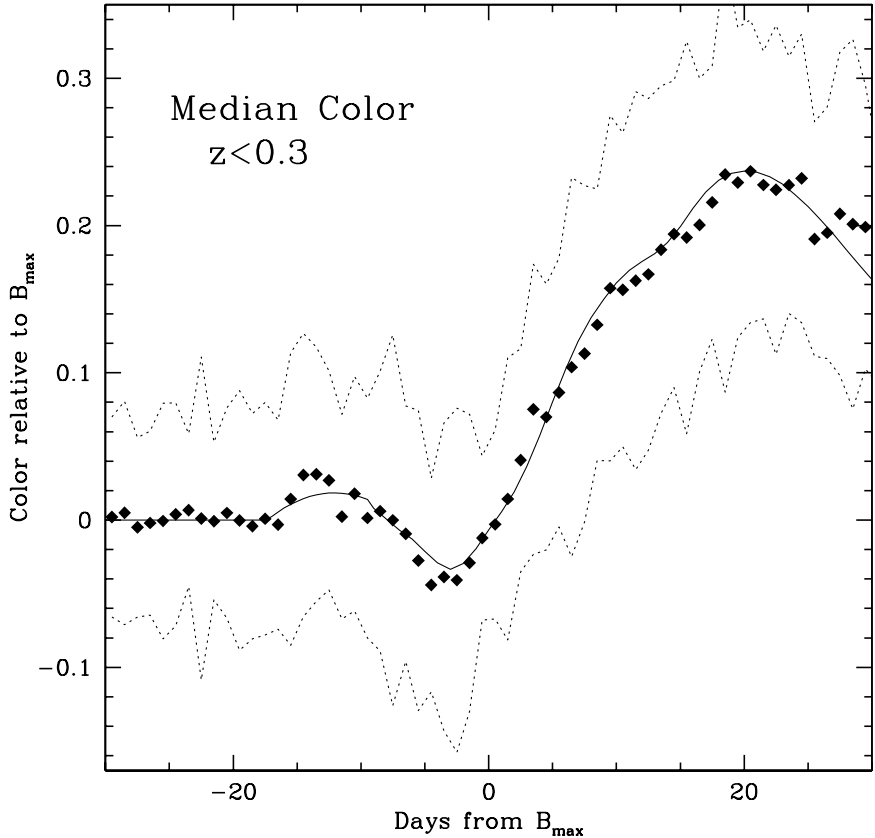}
\end{center}
\caption{ The $V-B$ flux color relative to B$_{max}$ as a function of SN phase. The data have been 
binned in 1-day (rest-frame) intervals and the median of each bin calculated. The diamonds represent 
the SDSS-II SNe with redshift less than 0.3. The solid line represents the color of the template. The 
dotted line is the 1$\sigma$ spread of the data. Calculating a color index for times later than -15 days, 
we find that the effective temperature increases linearly from -15 to -9 days then remains nearly constant 
until maximum. This is not expected from \citet{arnett82}, which predicts that the effective temperature 
should remain constant. This implies that the power-law index should be nearer to $n\approx 4$, rather 
than the $n\approx 2$ calculated from the observations. \label{color}}
\end{figure}

Figure~\ref{color} shows the flux color as a function of supernova phase. The colors have been binned 
by rest-frame day and the median calculated for each bin. As expected, the supernovae get significantly 
red after maximum as the $V$-band light curve fades more slowly than the $B$. Near the time of 
explosion, the color is slightly red and shifts to the blue just before maximum. This observed color 
change is exactly what is expected given the 2-stretch fit of the individual $B$- and $V$-band light curves 
(and this is shown by the solid line in the figure). Using the median $B$-band flux, we convert the flux 
color into magnitudes and find the color index at $-15$~days is $B-V=0.5$ which shifts to $B-V=0.0$ at 
$-9$~days. The data earlier than $-15$~days is too noisy to attempt estimating a reliable magnitude 
color index. We convert the color index to an effective temperature using the empirical relation from 
\citet{seki00} and find that $T_{eff}$ rises linearly from 6000 K at $-15$~days to 9500 K at 
$-9$~days and then stays nearly constant until maximum.

This linear rise in temperature over the first week after explosion is not expected from \citet{arnett82}, 
and is puzzling. If the color variation represents a temperature change and the optical bands are 
following the bolometric flux, then we expect a temporal power-law index of at least $n =4$ instead of 
the observed $n \approx 2$. In conclusion, we find that the power-law index for the SDSS SNe Ia agrees 
with the Arnett prediction; however, we find that the early light curves do not follow the prediction of slow 
and modest temperature changes.

\subsection{Source of Rise-Time Variation}

Our SDSS-II supernovae show a range of rise times for a fixed fall time, suggesting that the 
physics of the rise and fall epochs differ. \citet{arnett82} and \citet{pinto00} show that the light curves 
of SNe~Ia can be simply described by the deposition of energy from synthesized radioactive elements 
combined with the diffusion rate of energy out of the expanding nebula. Before maximum 
brightness the energy input rate from radioactivity exceeds the energy lost to luminosity. At 
maximum, the luminosity matches the instantaneous energy deposition rate and the decline from 
maximum correlates with the total radioactive yield. Using more detailed models, \citet{woos08} 
show that the shape of the light curve depends on more than the radioactive yield. For example, 
increasing the production of intermediate mass elements (e.g., silicon and calcium) at a fixed nickel 
mass ($^{56}$Ni) narrows the light curve while increasing the peak luminosity. This occurs because 
the total burned mass correlates with the KE of explosion, and the faster the 
supernova expands, the earlier the radioactive energy diffuses out. In contrast, a low KE 
means a slow rise and faint maximum. Note that the variation in KE trends in the 
opposite direction to the Phillip's relation. Observationally, the radioactive yield and KE 
influence the rise and fall times in different ways and variation in these two parameters could help 
explain the range of rise times seen in the SDSS-II supernovae. Other explosion parameters, such 
as non-radioactive iron yield or degree of mixing, could also impact the rise versus fall times 
\citep{woos08}. For illustration, we consider only KE variation here.

\begin{figure}[h!]
\begin{center}
\includegraphics*[scale=0.6]{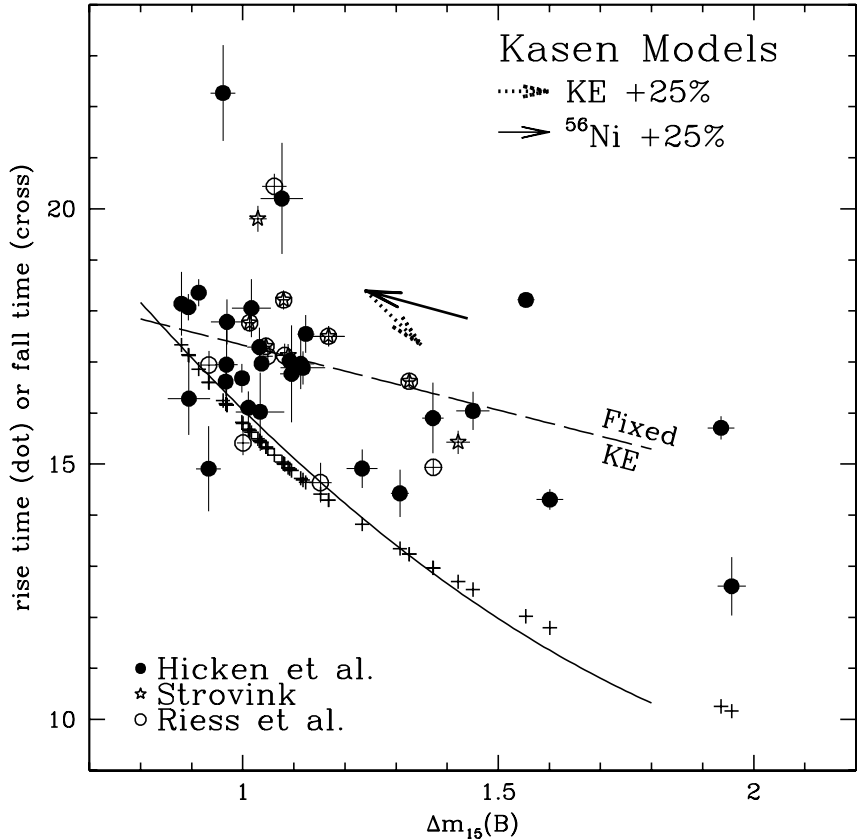}
\end{center}
\caption{Plot of rise time and fall time vs. \dm15 for the 14 nearby SNe~Ia analyzed in this study, 
with theoretical distributions derived from Kasen model light curves.  The dotted line is a 
representation of the Kasen model rise times and its shallow slope implies that rise time is only 
mildly dependent on radioactive yield. The solid line is a quadratic fit to the fall times estimated from 
the models and are very consistent with the observed fall times. The arrows represent how a 25\%\ 
change in KE or $^{56}$Ni yield would change the position of a specific SN~Ia from the Kasen 
models.  \label{kasen}}
\end{figure}

\citet{kasen07} have calculated model light curves for SNe~Ia with fixed KE and varying 
radioactive nickel masses to show that they can match the Phillip's relation fairly well. We can apply 
the 2-stretch method to these model curves almost as easily as to real data. From this we can
see if the rise and fall times of the models can refine the physics of the observed light curve shapes. 
Unfortunately, the MLCS2k2 fiducial does not match the model light curves sufficiently well to 
estimate accurate rise and fall times. Instead, we chose the $M_{Ni}=0.49$~M$_\sun$ model curve 
as a fiducial and applied the 2-stretch fitter to the remaining models. The resulting rise times for 
fixed KE models with varying nickel yield are shown as a dashed line in Figure~\ref{kasen} and 
compared with observed low-redshift rise/fall times. The model rise time varies only modestly with 
\dm15 (slope of $-2.6$ days/\dm15 ) compared to the fall time (slope of $-9.2$ days/\dm15 at 
\dm15=1.1). This is in stark contrast to the single-stretch parameterization where the rise time 
parallels the fall time in this diagram.

We compare the predicted effect on rise/fall times from varying $^{56}$Ni yield with the SDSS-II 
data in Figure~\ref{rise-fall}. The observed trend of $t_r-t_f$ increasing with larger \dm15 is well 
matched by the model, although there are still many supernovae in the sample that lie far from the 
model. Clearly, the single-stretch parameterization of light curve shape is not justified by the 
observed rise/fall-time variations or predictions of the $^{56}$Ni model. The small scatter in the rise 
portion of the SN~Ia light curves seen in the full SDSS-II sample (Figure~\ref{alldata}) should not be a 
surprise if it is a direct consequence of $^{56}$Ni yield being the dominant source of diversity in 
SNe~Ia \citep{arnett85}.  This is a popular interpretation for variations in peak brightness in SNe~Ia.

\citet{kasen07} models that vary KE at fixed nickel yield were also fit with the 2-stretch method. 
Arrows in Figure~\ref{kasen} illustrate the amplitude and direction in the rise/fall time versus \dm15 
plane of a supernova that increases its nickel yield by 25\% at fixed KE and then increases
its KE by 25\% at fixed radioactive yield. The change in KE produces a steeper variation in rise 
time than does nickel and in the opposite sense, i.e., increasing KE results in a faster light 
curve decay as well as a shorter rise time. Unlike nickel variations, changes in KE affect the rise 
time (slope of $-9.7$ days/\dm15) as strongly as the fall time, meaning pure variations in KE result
in almost no change in $t_r-t_f$ (and would be well fit by a single-stretch parameterization).

While this analysis focuses on the effects of KE and $^{56}$Ni yield, there are other 
parameters that may be important to the light curve shape as well. In particular, mixing the $^{56}$Ni to 
a larger radius may result in faster rise times \citep{pinto00b,hoef09}. This effect plays a secondary role 
\citep{hoef09} in light curve shape, however, and is not included in this analysis. The \citet{kasen07} 
models used in this analysis contain no effects from variations in $^{56}$Ni distribution. We have only 
employed model light curves with fixed iron mass yields, which is the parameter that controls the 
$^{56}$Ni distribution in the one-dimensional \citet{kasen07} models.

In principle, varying both $^{56}$Ni yield and KE will allow a supernova to reach any point on the 
$t_r-t_f$ versus \dm15 diagram (Figure~\ref{rise-fall}). Changes in radioactive yield will move 
supernovae diagonally, while KE variations shift supernovae horizontally. Given these motions, it 
is difficult to explain the handful of events with very slow rise times and normal decline rates
($t_r-t_f> 6$ and \dm15$\approx 1.0$). SN~1990N may be a member of this group in the low-redshift
set. The Kasen models also predict that peak luminosity should be related to nickel yield 
and KE variations, with an increase in $^{56}$Ni and KE resulting in brighter events.
\citet{woos08} found, however, that the effect of these parameters on light curve width works in the 
opposite sense. An increase in KE narrows the light curve while increasing peak luminosity and  a 
larger nickel mass widens the light curve while making a brighter peak. Examining the absolute 
magnitude estimates of the SDSS-II events may help sort out the origin of the rise/fall variations.

\subsection{Correlation with Host Galaxy Color}

\begin{figure}[h!]
\begin{center}
\includegraphics*[scale=0.6]{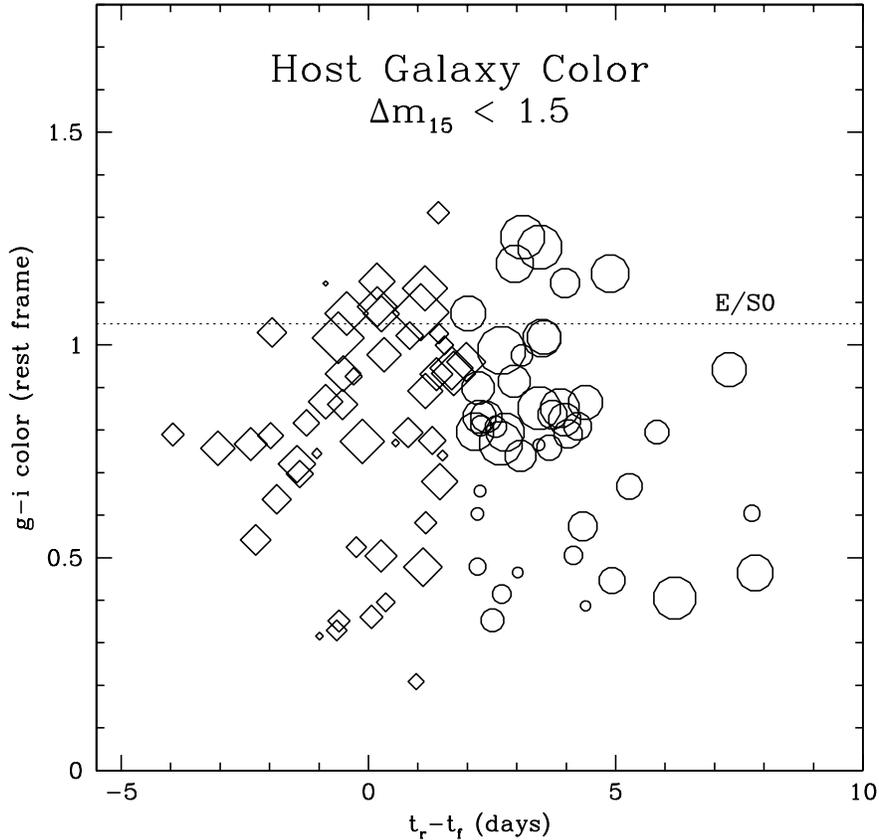}
\end{center}
\caption{Host galaxy color vs. the difference in rise and fall times. Galaxies that hosted SNe~Ia 
with $t_r-t_f<2.0$ days are plotted as diamonds while hosts with high RMF SNe are circles. The 
size of the plotting symbol correlates with the total brightness of the galaxy. The RMF cut at 2 days was 
selected arbitrarily. The host color distributions do not appear different between these two groups. 
\label{galaxy}}
\end{figure}

\citet{scan05} and \citet{sul06} have shown that rates of SNe Ia are strongly connected 
with host galaxy star formation rate. They model the rates as coming from two sources, one from a 
passive population of stars and the other a prompt population of supernovae correlated with high
star formation rates. It is natural to speculate that the range of  RMF corresponds to the 
sources of supernovae in the two stellar populations. To test this idea, we construct the $g-i$ color of 
the host galaxies of our sample of 105 SDSS-II supernovae. The $g-i$ color index correlates
well with specific star formation rate measured from the SDSS spectroscopic sample. The host 
magnitudes are taken from the SDSS-DR5 catalog and are color corrected to the rest frame by 
interpolating from tables in \citet{fuk96}. The host galaxy color versus supernova $t_r-t_f$ is shown 
in Figure~\ref{galaxy} for supernovae with \dm15$<1.5$ mag. Very fast declining supernovae are 
known to be associated with early-type galaxies (e.g., \citet{gal05, hamuy96a}) and might confuse 
the result.

There is no significant correlation between host $g-i$ color and the rise-time properties of the 
supernovae. Supernovae with RMF $< 2.0$ days tend to have slower decline rates than the other 
supernovae and can be expected to result from the prompt, high star-formation rate population. We 
do not observe this relation in our sample. 

\subsection{Correlation with Hubble Residuals}

The Phillip's relation shows that the SN~Ia light curve decline rate is related to the optical peak 
luminosity. The SDSS-II light curves suggest that there are a range of basic light curve shapes and 
this leads to several questions:

\begin{itemize}

\item Which parameter is the best indicator of peak luminosity: fall time, rise time,
total width, or rise minus fall time?

\item Is there a difference in the average peak luminosity between fast and slow risers
for a fixed decline rate?

\item Do the color or dust properties vary with rise time?

\end{itemize}

To investigate these questions, we estimated the apparent peak $V$-band magnitude, and peak 
color by fitting the 105 light curves with good rise/fall-time measurements with MLCS2k2. We 
relaxed all priors to the fits (flatnegav) and assumed a fixed extinction law of $R_V=2.5$. Despite 
the good rise-time information in the light curve data, we restricted the fit to include only those 
points later than $-10$~days before maximum; this was done because we found rather poor fits to 
the light curves when using the MLCS2k2 template extrapolations that put the explosion date at 
$-20$ days before $B$ maximum. For this analysis, we cut nine supernovae with \dm15$>1.5$ 
because their colors are strongly dependent on decline rate \citep{gar04}.

We also fit the light curves using the SALT-II software \citep{salt2} and compared the color 
parameter, $c$, and peak apparent magnitudes to those from MLCS2k2. The average color 
difference was $(B-V)-c=-0.027$ with an rms dispersion of 0.045 mag and the average apparent 
magnitude difference was $m_V$(mlcs)$-m_V$(salt)$=0.019$ with an rms dispersion of 0.038 mag. 
We conclude that for these well-sampled light curves the differences in fit parameters between 
MLCS2k2 and SALT-II are insignificant.

Given the estimated apparent peak magnitude, redshift, $(B-V)$ color, and our measured rise/fall 
stretches, we minimized the residuals on the Hubble diagram calculated from

$$\Delta m = m_V-5log(d_L)+25-\alpha +\beta (B-V)+\gamma (s_f-1)+\delta (s_r-1)+\epsilon
(t_r-t_f) $$,

where $\alpha$, $\beta$, $\gamma$, $\delta$, and $\epsilon$ are coefficients allowed to vary in the 
minimization. The luminosity distance ($d_L$ in Mpc) is calculated from the known redshift
and assuming the cosmological parameters of $H_0=65$~km~s$^{-1}$~Mpc$^{-1}$, $\Omega_
\Lambda=0.7$, and $\Omega_m=0.3$. Choosing a different set of cosmological parameters changes the
relative absolute magnitudes by only a few percent at these redshifts. An `amoeba' algorithm was 
employed for the minimization of the $\chi^2$ parameter, $$\chi^2 = \sum_{i}{\Delta m_i^2/\delta
\mu_i^2}$$, where $\delta\mu_i$ is the distance modulus uncertainty from the MLCS2k2 fit. We calculate 
$\chi^2$ for several combinations of the free parameters and the results of these fits are shown in 
Table~\ref{tbl-2}.

\begin{deluxetable}{lccccccc}
\tabletypesize{\scriptsize}
\tablecaption{Hubble Residuals \label{tbl-2}}
\tablewidth{0pt}
\tablehead{
\colhead{Fit}       & \colhead{96 SNe} & \colhead{96 SNe} & \colhead{96 SNe} & \colhead{96 SNe} & \colhead{96 SNe} & \colhead{52 SNe (t$_r-$t$_f<2.0$)} & \colhead{44 SNe (t$_r-$t$_f>2.0$)} \\
\colhead{Parameters}& \colhead{ 1 }    & \colhead{ 2 }    & \colhead{ 3 }    & \colhead{ 4 }    & \colhead{ 3 }    & \colhead{ 3}    & \colhead{3}
}
\startdata
$M_V$ ($\alpha$)    & $-19.087$ & $-19.093$ & $-19.067$ & $-19.066$ & $-19.061$ & $-19.054$ & $-19.070$ \\  
$(B-V)$ ($\beta$)      & ...       & $-1.466$  & $-1.236$  & $-1.251$  & $-1.213$  & $-1.343$  & $-1.082$  \\
$s_f-1$ ($\gamma$)     & ...       & ...       & 0.614     & 0.583     & ...       & ...     & ...     \\
$s_r-1$ ($\delta$)     & ...       & ...       & ...       & 0.792     & ...       & ...       & ...       \\
$s_r+s_f-2.0$ ($\epsilon$) & ...   & ...       & ...       & ...       & 0.691     & 0.600       & 0.765       \\
\tableline
$\sigma$            & 0.206     & 0.175     & 0.163     &  0.145    & 0.145     & 0.143     & 0.145     \\
$\chi^2$/dof        & 214/95    & 147/94    & 131/93    & 100/92    & 101/93    & 59/49      & 42/41      \\
\enddata
\tablecomments{Residual = m$_V-$5log($d_L$)+25$-\alpha$+$\beta$(B$-$V)+$\gamma$(s$_f-$1)+$\delta$(s$_r-$1)+$\epsilon$
(t$_r$+t$_f$)}
\end{deluxetable}

The scatter about the Hubble line for a one parameter fit (zero point) applied to the remaining 96 
events provides an rms scatter of 0.21~mag. This small scatter is not surprising given that the 
quality of the data is excellent and we have eliminated intrinsically red supernovae with the 
\dm15$>1.5$ cut.

Adding a color term reduces the scatter to 0.18~mag. The size of the color term, $
\beta=-1.47$, should correspond to $R_V$, the ratio between $V$-band extinction and reddening in 
$B-V$, if dust were the source of the color variation. As noted for both local and high-redshift 
supernovae (e.g., \citet{kessler08}), the best-fit color term tends to be much smaller than the 
standard Milky Way extinction value of $R_V=3.1$. This may indicate non-standard dust in other 
galaxies or that the supernova colors are not well understood.

Minimizing the scatter by adding the fall stretch, $s_f$, as a third parameter has some effect on the 
Hubble residuals, reducing the scatter to 0.16~mag. A fourth parameter that includes the rise 
stretch, $s_r$, or the rise minus fall time, $t_r-t_f$, provides the smallest scatter of 0.145~mag. 
Surprisingly, the Hubble residuals are minimized when both the rise and fall stretches have large,
positive coefficients. The Kasen models that vary both nickel yield and KE predict that increasing 
rise time (lowering KE) should reduce the supernova peak brightness, so the rise-time parameter 
would have the opposite sign from the fall-time parameter. Our hypothesis that  KE has a strong 
influence on light curve shape and brightness does not match the observations well, but other 
variables should be explored. Varying the mass of stable iron elements appears to affect the rise 
time and brightness in the same directions as the KE variation so are also disfavored by this 
analysis. 

Applying a three-parameter minimization with the light curve characterized by $t_r+t_f$ results in 
the same scatter as a four-parameter fit with $t_r$ and $t_f$ as separate light curve shape 
indicators. This result suggests that it is the total width of the light curve that correlates with peak 
luminosity and not the rise or fall separately. Indeed, a three-parameter fit with $t_r-t_f$ does no 
better in reducing the scatter than a two-parameter minimization with just zero point and color.

\begin{figure}[h!]
\begin{center}
\includegraphics*[scale=0.6]{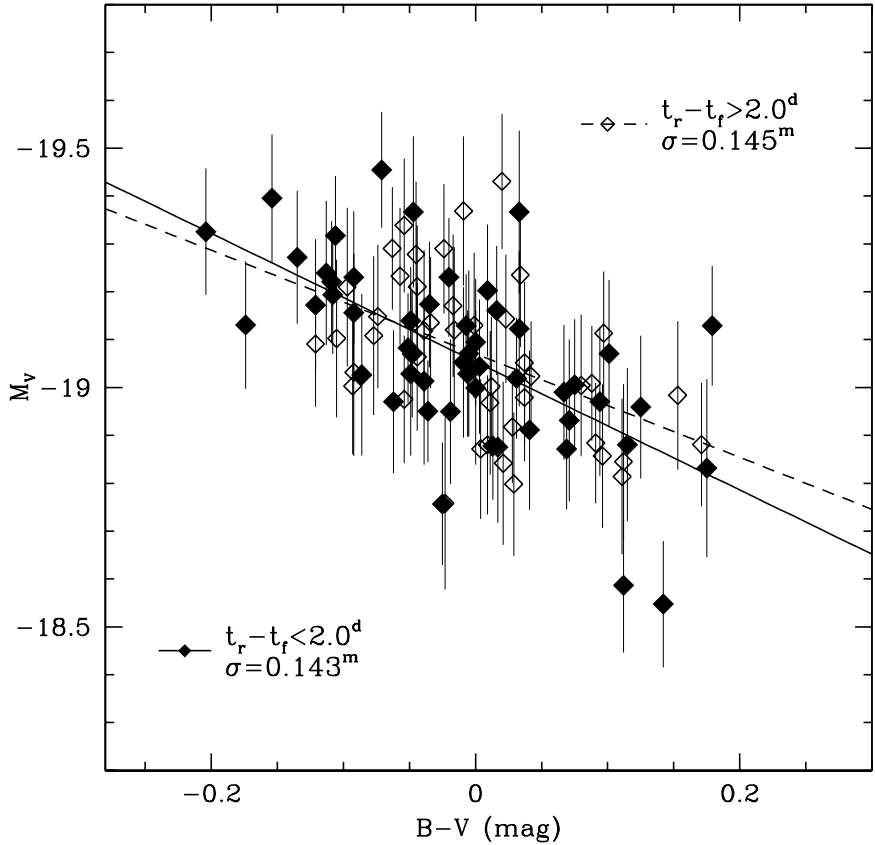}
\end{center}
\caption{Absolute $V$-band magnitude vs. peak color for fast-rising ($t_r-t_f<2.0$ days; 
filled diamonds) and slow-rising ($t_r-t_f>2.0$ days; open diamonds) SDSS-II supernovae after 
correction for light curve shape. No difference in peak luminosity or color slope is apparent. Note, 
however, that the fast-rising events tend to be more blue at peak with 9 of the bluest 10 supernovae 
having $t_r-t_f<2.0$.
\label{hubble1}}
\end{figure}

When we divide the sample into a `low RMF' group with $t_r-t_f<2.0$~days and a `high RMF' 
subset with $t_r-t_f>2.0$~days there appears to be no substantial differences in the Hubble 
residuals. Applying a three-parameter minimization (zero point, color, and total stretch), we find the color 
term for the low and high RMF is only marginally different (see Figure~\ref{hubble1}). We note, 
however, that the low RMF group has more blue supernovae than the high RMF group, with 9 of 
the 10 bluest events having $t_r-t_f<2.0$~days. The relation between luminosity and full width of 
the light curve also shows a little difference between the two rise-time divisions.

The average absolute magnitudes of the two groups after color and light curve shape correction 
differ by less than 0.05~mag, implying that the presence of a range in rise times  has very little direct 
impact on cosmological measurements.

\subsection{Estimating $^{56}$Ni Yields}

Arnett's rule \citep{arnett82} states that the bolometric luminosity of a SN~Ia at maximum light is very 
close to the instantaneous energy being deposited by the synthesized radioactive elements. This 
equality has been used by several studies (e.g., \citet{stritz05}; \citet{cont00}; \citet{how08}) to 
estimate the mass of $^{56}$Ni created in the supernova. However, the time between explosion and 
maximum light is a key parameter in this calculation and it has often been assumed to be a constant 
19.5~days. In this paper, we have found that the average rise time is shorter than that estimated in 
previous studies and it is not strictly linked to the decline speed. This will change the distribution of $
^{56}$Ni yields when compared with earlier assumptions about rise-time properties.
We estimate $^{56}$Ni yields using the prescription of \citet{stritz05} but with a simplified technique 
for measuring the bolometric luminosity. We used the time from explosion to peak flux as our 
measured rise-time values. We have already calculated the extinction-corrected absolute $V$-band 
magnitudes for the well-observed objects in the SDSS-II sample and we use these to scale an 
average type~Ia spectrum which we integrate over a wide wavelength range to get quasi-bolometric 
$UVOIR$ luminosity. For a supernova with absolute magnitude $M_V$=0.00 (Vega system) and the 
\citet{hsiao07} spectral template at $B$-band maximum, we find a bolometric flux of 2.04$\times 
10^{-5}$ erg~cm$^{-2}$~s$^{-1}$ between 300 and 1000~nm. This approximation is best for normal 
type~Ia and is not as accurate for sub-luminous events which tend to be more red than normal at 
maximum. However, \citet{cont00} found that the $V$ band is an excellent indicator of bolometric 
energy even at late phases when the spectrum has become red. Under these assumptions, the 
$^{56}$Ni yield is simply $$M_{Ni}\; =\; {{4\pi \; 9.52\times 10^{38}\; 10^{-M_V/2.5}\; 2.04\times 
10^{-5}}\over{\alpha \; 6.45\times10^{43}\exp{(-t_r/8.8)}+1.45\times 10^{43}\exp{(-t_r/111)}}}
\;\;\;\;\;\;\;\;\  (M_\odot) $$,
where $t_r$ is the rise time in days and $\alpha$ is a constant describing the accuracy of Arnett's 
rule (we assume $\alpha=1$).

\begin{figure}[h!]
\begin{center}
\includegraphics*[scale=0.6]{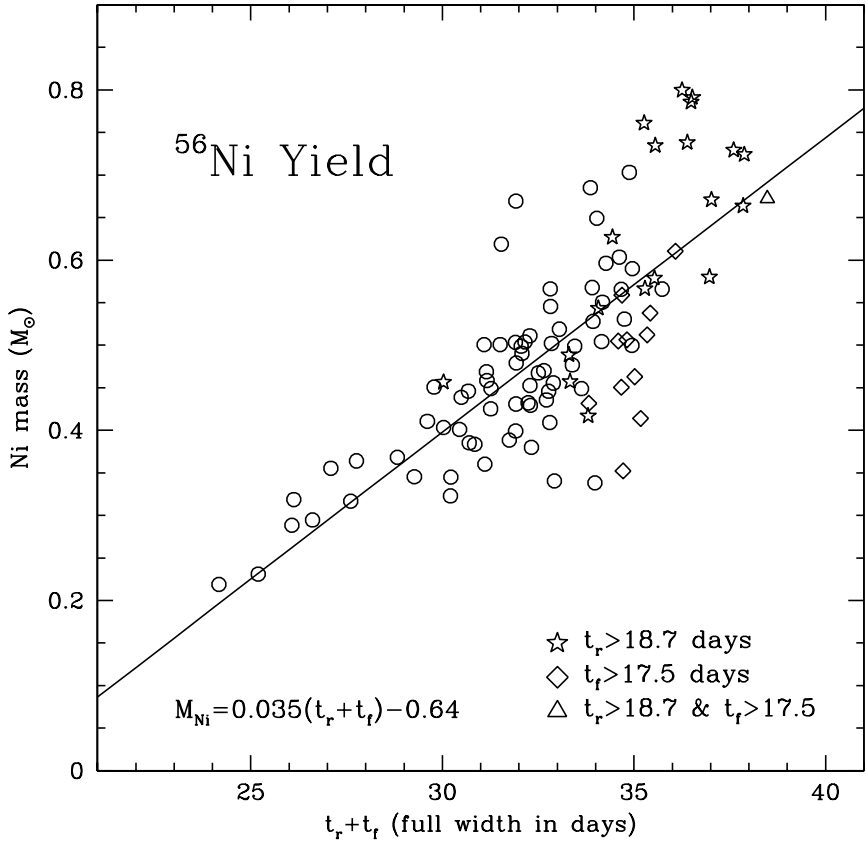}
\end{center}
\caption{Plot of the derived $^{56}$Ni mass for the SDSS-II
supernovae vs. the full width (rise plus fall times)
of the $B$-band light curve. The $^{56}$Ni mass
is estimated from Arnett's rule so it is a function of
both the peak bolometric luminosity and the rise time.
The solid line is a least-squares fit to the points.
Supernovae with slow rise times ($t_r>18.7$ days) are
plotted as stars while those with slow declines
($t_f>17.5$ days) are shown as diamonds. Only one
light curve had both a slow rise and decline and it
is plotted as a triangle with a full width of 38 days.
The supernovae with the largest $^{56}$Ni
yields ($>0.7 M_\odot$) have slowly rising light curves
while slowly fading events produce more typical
amounts ranging between 0.4 and 0.6 M$_\odot$
of $^{56}$Ni. Clearly, knowledge of the rise time is critical in
understanding the physics of type Ia events.
\label{nimass}}
\end{figure}

The estimated $^{56}$Ni yields for the SDSS-II events range from 0.2 to 0.8 $M_\odot$, which are 
typical for this method \citep{stritz06}. Figure~\ref{nimass} shows the $^{56}$Ni yield as a function of 
the total light curve width ($t_r+t_f$) and there is a strong correlation between these two quantities. 
Most of the supernovae producing the largest amount of $^{56}$Ni are events with $t_r>18.7$~days 
(slowest rising 20\%). In contrast, the slowest declining supernovae ($t_f>17.5$~days) produce 
average yields of $^{56}$Ni. This demonstrates the importance of independently measuring the rise 
time of SN~Ia to understand the physics of the thermonuclear explosion mechanisms.

\clearpage

\section{Conclusion}

The SDSS-II supernova sample provides tight constraints on the rise time and shape of SN~Ia light 
curves. The $k$-corrected, median light curves of 391 events shows a $B$-band rise time of 16.8 days 
and a significantly smaller dispersion on the rise portion of the curve than on the fading side. This 
implies that the rise time is less impacted by variations in radioactive nickel yield than the fall, as 
predicted by the \citet{kasen07} model light curves. It is clear that the single-stretch parameter 
commonly used to characterize SN~Ia light curves is not capable of capturing the full range of light 
curve shapes, especially when pre-maximum observations are available.

We selected 105 SNe~Ia from the larger sample that had sufficient photometric precision and 
cadence to yield rise- and fall-time errors of less than 2.0 days. From this set, we find the rise time in 
the $B$ band is $t_r=17.38\pm 0.17$ days (standard error of the mean). This is significantly 
different from the rise time of 19.5~days measured from local supernovae \citep{rie99} and high-redshift
events \citep{con06}. We find the cause of this discrepancy to be the difference in the 
fiducial curves used in the analyses and our application of a 2-stretch fitting method that permits the 
rise and fall to be fit independently. We find rise times ranging from 13~days to 23~days for events 
with ``normal" decline times of around 15~days, demonstrating that the rise and fall are not strictly 
correlated. 

We applied our 2-stretch light curve fitting method to data from SDSS-II to test the conjecture by 
\citet{str07} of a bimodal distribution in SN~Ia light curve shape. From our 105 high-quality light 
curves, we are not able to reproduce this bimodal distribution in the difference between rise and fall 
times.  We do find a significant spread in RMF times for the sample, which implies again that the rise 
and fall are not strictly correlated. This large range in RMF is expected from explosion models that
simply vary the radioactive nickel yield.

Contrary to the premise of the single-stretch method, many of the slowest declining SNe are among 
the fastest risers. This effect is better modeled through MLCS2k2, as shown by our simulations, and 
this may be a result of MLCS2k2 being trained on a large number of real nearby SNe Ia. While 
MLCS2k2 is better than single stretch at representing the observed light curves, the simulations 
with MLCS2k2 do not capture the full spread in rise times evident for normal-declining SNe Ia. 
Specifically our simulations using MLCS2K2 do not contain normal-declining events that are slow 
rising.

We investigated the relation between peak luminosity, light curve shape, and color by minimizing 
the scatter on the Hubble diagram for the 96 SNe~Ia with good rise/fall data and \dm15$<1.5$ . The 
Hubble residuals have the smallest scatter when using both rise and fall time as fit parameters. The 
correlation of rise and fall time with luminosity have the same sign, meaning that the full width of the 
light curve is the best indicator of luminosity. Twelve models suggested that a second 
parameter, such as KE, competing with radioactive nickel yield, could be revealed in the 
rise and fall times and their correlation with luminosity. However, rise and fall times are found to act 
in concert in the SDSS-II SNe, implying that KE (or other physical parameters that 
broaden the light curve and lower peak luminosity) is not a major contributor to light curve shape.

\citet{pignata08} recently suggested that SNe~Ia with high Si II velocity gradients (HVG) may have fast- rising light curves. We are analyzing the spectra of the SDSS sample to see if there is a rise-time, 
spectral velocity gradient correlation. It is interesting to note that SN~1990N, a low-redshift 
supernova with a very slow rise, has a low Si II velocity gradient, but a more complete sample needs 
to be constructed. 

The application of the 2-stretch fitting method we have developed provides a better fit to SN~Ia 
light curves than simply stretching the entire time axis. By independently fitting the rise and
fall portions of each light curve, we found that the
rise and fall times are not strictly correlated. When correcting SN~Ia luminosity
for light curve shape, adding the rise-time parameter to the fall
time reduces the scatter in the Hubble diagram when compared to using the fall time alone.
However, we find that the full width of the light curve (rise plus fall time) minimizes the scatter about the
Hubble line as well as using both rise and fall times as independent parameters. This does
suggest some danger doing cosmology by combining a low-redshift supernova set dominated by poor pre-maximum sampling with a well-sampled high-redshift set. 

Our results show that measurement of rise time as a separate 
parameter characterizing SN~Ia light curve shape may be a useful diagnostic in understanding the 
progenitors and explosion mechanisms of thermonuclear events. Based on $^{56}$Ni yield calculations 
in \citet{cont00} and \citet{stritz05}, our average rise time would result in a 15\%-20\% reduction in 
$^{56}$Ni yield as compared to models that use a 19.5-day rise time. However, there are also some 
SNe Ia that have rise times longer than 19.5 days, resulting in a higher $^{56}$Ni yield. It is clear that
accurate nickel yield estimates require a good measurement of the rise time for individual
events. Future supernova searches that provide well-sampled, high-quality data well before maximum
light will reveal more about the nature of SN~Ia.

\acknowledgements
Funding for the creation and distribution of the SDSS and SDSS-II
has been provided by the Alfred P. Sloan Foundation,
the Participating Institutions,
the National Science Foundation,
the U.S. Department of Energy,
the National Aeronautics and Space Administration,
the Japanese Monbukagakusho,
the Max Planck Society, and the Higher Education Funding Council for England.
The SDSS Web site \hbox{is {\tt http://www.sdss.org/}.}

The SDSS is managed by the Astrophysical Research Consortium
for the Participating Institutions.  The Participating Institutions are
the American Museum of Natural History,
Astrophysical Institute Potsdam,
University of Basel,
Cambridge University,
Case Western Reserve University,
University of Chicago,
Drexel University,
Fermilab,
the Institute for Advanced Study,
the Japan Participation Group,
Johns Hopkins University,
the Joint Institute for Nuclear Astrophysics,
the Kavli Institute for Particle Astrophysics and Cosmology,
the Korean Scientist Group,
the Chinese Academy of Sciences (LAMOST),
Los Alamos National Laboratory,
the Max-Planck-Institute for Astronomy (MPA),
the Max-Planck-Institute for Astrophysics (MPiA), 
New Mexico State University, 
Ohio State University,
University of Pittsburgh,
University of Portsmouth,
Princeton University,
the United States Naval Observatory,
and the University of Washington.

This work is based in part on observations made at the 
following telescopes.
The Hobby-Eberly Telescope (HET) is a joint project of the University of Texas
at Austin,
the Pennsylvania State University,  Stanford University,
Ludwig-Maximillians-Universit\"at M\"unchen, and Georg-August-Universit\"at
G\"ottingen.  The HET is named in honor of its principal benefactors,
William P. Hobby and Robert E. Eberly.  The Marcario Low-Resolution
Spectrograph is named for Mike Marcario of High Lonesome Optics, who
fabricated several optical elements 
for the instrument but died before its completion;
it is a joint project of the Hobby-Eberly Telescope partnership and the
Instituto de Astronom\'{\i}a de la Universidad Nacional Aut\'onoma de M\'exico.
The Apache 
Point Observatory 3.5 m telescope is owned and operated by 
the Astrophysical Research Consortium. We thank the observatory 
director, Suzanne Hawley, and site manager, Bruce Gillespie, for 
their support of this project.
The Subaru Telescope is operated by the National 
Astronomical Observatory of Japan. The William Herschel 
Telescope is operated by the 
Isaac Newton Group, and the Nordic Optical Telescope is 
operated jointly by Denmark, Finland, Iceland, Norway, 
and Sweden, both on the island of La Palma
in the Spanish Observatorio del Roque 
de los Muchachos of the Instituto de Astrofisica de 
Canarias. Observations at the ESO New Technology Telescope at La Silla
Observatory were made under program IDs 77.A-0437, 78.A-0325, and 
79.A-0715.
Kitt Peak National Observatory, National Optical 
Astronomy Observatory, is operated by the Association of 
Universities for Research in Astronomy, Inc. (AURA) under 
cooperative agreement with the National Science Foundation. 
The WIYN Observatory is a joint facility of the University of 
Wisconsin-Madison, Indiana University, Yale University, and 
the National Optical Astronomy Observatories.
The W. M. Keck Observatory is operated as a scientific partnership 
among the California Institute of Technology, the University of 
California, and the National Aeronautics and Space Administration. The 
Observatory was made possible by the generous financial support of the 
W. M. Keck Foundation. 
The South African Large Telescope of the South African Astronomical 
Observatory is operated by a partnership between the National 
Research Foundation of South Africa, Nicolaus Copernicus Astronomical 
Center of the Polish Academy of Sciences, the Hobby-Eberly Telescope 
Board, Rutgers University, Georg-August-Universit\"at G\"ottingen, 
University of Wisconsin-Madison, University of Canterbury, University 
of North Carolina-Chapel Hill, Dartmouth College, Carnegie Mellon 
University, and the United Kingdom SALT consortium. The Italian Telescopio Nazionale Galileo is 
operated on the island of La Palma by the Fundacion Galileo Galilei of the Instituto Nazionale
di Astrofisica at the Spanish Observatorio del Roque de los Muchachos
of the Instituto de Astrofisica de Canarias. 

DARK is funded by the Danish National Research Foundation. The Oskar Klein Centre is funded by The 
Swedish Research Council. Jesper Sollerman is a Royal Academy of Sciences Research Fellow 
supported by a grant from the Knut and Alice Wallenberg Foundation. This research is supported at
Rutgers University by the US Department of Energy grant DE-FG02-08ER41562 (PI: Jha).
This work is partially supported by U.S. National Science Foundation through grant AST-0507475 (ESSENCE). 

We also thank Dr. Peter Nugent for many helpful and insightful comments.

\end{document}